\documentclass[aps,twocolumn,superscriptaddress]{revtex4-1}
\usepackage{multirow}
\usepackage{graphicx}
\usepackage{longtable}
\usepackage[utf8]{inputenc}
\usepackage[T1]{fontenc}
\usepackage{epstopdf}
\usepackage{textcomp}
\usepackage[usenames,dvipsnames]{color}
\usepackage{amsbsy}
\usepackage{amsmath}
\usepackage{amssymb}
\usepackage{amsfonts}
\usepackage{dsfont}
\usepackage{mathtools}
\usepackage{braket}
\usepackage{array}
\usepackage{gensymb}
\usepackage{placeins}
\usepackage{dashrule}
\usepackage{xcolor}
\usepackage{booktabs}
\usepackage{hyperref}

\setcitestyle{super}

\newcommand{\ra}[1]{\renewcommand{\arraystretch}{#1}}

\usepackage[normalem]{ulem}

\newcommand{\dH}{\text{\itshape\DH}}
\newcommand{\vcn}[1]{\hat{#1}}
\newcommand{\vcc}[1]{{\vec{#1}}}
\newcommand{\vc}[1]{\boldsymbol{\mathrm #1}}
\newcommand{\matcc}[1]{\overset{\leftrightarrow}{#1}}
\newcommand{\matt}[1]{\mathcal{#1}} 
\newcommand{\mat}[1]{\underline{\underline{#1}}}
\renewcommand{\Im}{\mathrm{Im}}
\renewcommand{\Re}{\mathrm{Re}}
\newcommand{\expo}[1]{\ensuremath{\mathrm{e}^{#1}}}
\newcommand{\abs}[1]{\lvert#1\rvert}                  
\DeclareMathOperator{\sign}{sign}
\makeatletter
\newcommand\sh@ftsym[1]{%
  \smash{\raise-.3ex\hbox{$\scriptscriptstyle#1$}}}
\newcommand\Gtrless{\mathbin{\sh@ftsym({\gtrless}\sh@ftsym)}}
\makeatother
\newcommand{\eg}{e.g.,\@ }
\newcommand{\ie}{i.e.\@ }
\newcommand{\cf}{cf.\@ }
\newcommand{\etal}{\textit{et al.\@ }}
\newcommand{\abinitio}{\textit{ab initio}}

\renewcommand\figurename{{\bf Figure}}
\renewcommand\tablename{{\bf Table}}
\renewcommand{\thefigure}{{\bf\arabic{figure}}}
\renewcommand{\thetable}{{\bf\arabic{figure}}}
\renewcommand{\onlinecite}[1]{ref.~\nocite{#1}\citenum{#1}} 

\newtheorem{theorem}{Theorem}

\newcommand{\R}{\mathbb{R}}
\newcommand{\St}{\mathbb{S}^2}
\newcommand{\m}{\boldsymbol{m}}
\newcommand{\del}{\partial}
\newcommand{\ein}{\hat{\vc{e}}}
\newcommand{\bs}[1]{ \vc{#1}}

\makeatletter
  \def\my@tag@font{\normalsize}
  \def\maketag@@@#1{\hbox{\m@th\normalfont\my@tag@font#1}}
  \let\amsmath@eqref\eqref
  \renewcommand\eqref[1]{{\let\my@tag@font\relax\amsmath@eqref{#1}}}
\makeatother

 \allowdisplaybreaks

\begin{document}
\title{
Antiskyrmions stabilized at interfaces by anisotropic Dzyaloshinskii-Moriya interaction}

\newcommand{\fz}{Peter Gr\"unberg Institut and Institute for Advanced Simulation, Forschungszentrum J\"ulich and JARA, 52425 J\"ulich, Germany}
\newcommand{\iceland}{Science Institute of the University of Iceland, VR-III, 107 Reykjav\'{i}k, Iceland}
\newcommand{\rwth}{Department of Mathematics I \& JARA FIT,
RWTH Aachen University, 52056 Aachen, Germany}

\author{Markus Hoffmann}
\email{m.hoffmann@fz-juelich.de}
\affiliation{\fz}
\author{Bernd Zimmermann}
\affiliation{\fz}
\author{Gideon P. M\"uller}
\affiliation{\fz}
\affiliation{\iceland}
\author{Daniel Sch\"urhoff}
\affiliation{\fz}
\author{Nikolai S. Kiselev}
\affiliation{\fz}
\author{Christof Melcher}
\affiliation{\rwth}
\author{Stefan Bl\"ugel}
\affiliation{\fz}

\date{\today}

\begin{abstract}
Chiral magnets are an emerging class of topological matter harbouring localized and topologically protected vortex-like magnetic textures called skyrmions, which are currently under intense scrutiny as a new entity for information storage and processing. Here, on the level of micromagnetics we rigorously show that chiral magnets can not only host skyrmions but also antiskyrmions as least-energy configurations over all non-trivial homotopy classes. We derive practical criteria for their occurrence and coexistence with skyrmions
that can be fulfilled by (110)-oriented interfaces in dependence on the electronic structure. Relating the electronic structure to an atomistic spin-lattice model by means of density-functional calculations and minimizing the energy 
on a mesoscopic scale 
applying spin-relaxation methods,
we propose a double layer of Fe grown on a W(110) substrate as a practical example. We conjecture that ultrathin magnetic films grown on semiconductor or heavy metal substrates with $C_{2v}$ symmetry are prototype classes of materials hosting magnetic antiskyrmions.
\end{abstract}
\maketitle
Chiral magnetic skyrmions~\cite{Bogdanov_89} are currently subject of intense scientific investigations as they are part of the topological revolution that is currently taking place in condensed matter, with the real-space topology of their magnetic structure playing a leading role. The topological protection of their magnetic structure and particle-like properties with a well-defined topological charge~\cite{footnote1} of $Q=-1$ offer good conditions for skyrmions becoming the new information-carrying particles in the field of spintronics~\cite{Fert_13}, and this explains the additional motivation of their current intensive investigation.\par

Skyrmions in chiral magnets may appear as isolated solitons or condensed in regular lattices. Their stability results from the Dzyaloshinskii-Moriya interaction (DMI)~\cite{Dzyaloshinskii, Moriya}, which breaks the chiral symmetry of the magnetic structure. The energy and size is determined by the competition between the Heisenberg, Dzyaloshinskii-Moriya, and Zeeman interaction together with the magnetic anisotropy energy (MAE), here expressed in terms of the spin-lattice model applied to an interface geometry

\begin{align}
H = &-\sum_{ij}J_{ij} \left( \vc{S}_i \cdot \vc{S}_j \right)\nonumber -\sum_{ij}\vc{D}_{ij} \cdot \left( \vc{S}_i \times \vc{S}_j \right)\nonumber\\
&- \sum_{i} \vc{B} \cdot \vc{S}_i + \sum_{i} K_{\perp} (\vc{S}_i \cdot \hat{\vc{e}}_z)^2\, 
\label{eq:ham}
\end{align}
with classical spin $\vc{S}$ of length one at atomic sites $i,j$ and the corresponding microscopic pair $J_{ij}$, $\vc{D}_{ij}$ and on-site $K_{\perp}$ interactions. The DMI results from the spin-orbit interaction and is only non-zero for solids lacking bulk ($\vc{r}\nrightarrow -\vc{r}$) or structure inversion symmetry ($z\nrightarrow -z$).
For applications in spintronics \cite{Kiselev_11, Fert_13, Nagaosa_13} skyrmions stabilized in systems with surface or interface induced DMI seem to be more promising than bulk systems with DMI. For these systems the dipolar interaction between magnetic moments is typically of minor importance and is added as an additional on-site contribution to the magnetic anisotropy energy of the interface, $K_{\perp}$ ($K_{\perp}<0$ means, the easy axis of the magnetization is normal to the interface plane, \ie along $\hat{\vc{e}}_z$ direction). 

The first example of a topologically non-trivial structure stabilized by the interface DMI was revealed for an Fe monolayer on an Ir(111) substrate \cite{Heinze_11}. Subsequent incremental modifications e.g.\ adding a monolayer Pd on top of the Fe layer, led to the observation of single skyrmions stablized by interface DMI~\cite{Romming_13,Dupe_14}. 
With respect to future applications, interfaces offer a great variety of options for optimizing and controlling magnetic parameters: Variation of the interface composition \cite{Heide_09}, of the interface crystal symmetry \cite{Bergmann_15}, the film thickness \cite{Romming_13}, as well as the fabrication of interlayers \cite{Nandy_16} and multilayers \cite{Dupe_16}.

Taking the micromagnetic view of the DMI, which is valid in the limit of slowly varying magnetic textures, with the prototypical examples of interfacial or cubic DMI,
\begin{equation} \label{eq:proto}
D   \left[ \m (\nabla \cdot \m) - (\m \cdot \nabla) \m \right]_z \quad \text{or} \quad 
 D \, \m\cdot (\nabla\times\m), 
\end{equation}
respectively, for $\m=\m(\vc{r})$ with $\vc{m}(\vc{R}_i)=\vc{S}_i$, it may not be surprising that the community focuses primarily on the realization of skyrmions ($Q=-1$) rather than \textit{antiskyrmions} characterized by the topological charge $Q=+1$. The classical Skyrme problem exhibits a reflection symmetry, and the particle and antiparticle with the topological charges $Q=\pm 1$ exist with the same minimal energy \cite{Piette_95, Lin_04, Manton_14}. This is different for chiral magnetic skyrmions, where this degeneracy is lifted as a consequence of chiral symmetry breaking. In the context of
cubic DMI it has been shown that corresponding antiskyrmionic configurations have a strictly higher energy~\cite{Melcher_14}. In fact, the densities in \eqref{eq:proto} can be mapped onto each other by a rigid ($90$ degree) rotation in horizontal spin space and can be considered equivalent.
A close inspection of those proofs and simulations which verify that the lowest-energy magnetization configuration of non-trivial topology is attained for $Q=-1$, reveals that a micromagnetic DMI Hamiltonian of the form \eqref{eq:proto} is assumed with a scalar DM constant $D$, also coined the spiralization.

Different versions of the DMI arise from rigid $O(2)$ transformations applied to the cubic DMI in \eqref{eq:proto}. Some of which, e.g.\ those transforming to bulk crystals with $S_4$ or $D_{2d}$ symmetry, can lead to the stabilization of antiskyrmions rather than skyrmions, as has been recently demonstrated for an acentric tetragonal MnPtPdSn Heusler compound~\cite{arXiv:1703.01017}
Due to symmetry restrictions, those $O(2)$ transformations are not available for the interfacial DMI \eqref{eq:proto} and thus antiskyrmions are absent in two-dimensional (2D)  or film systems described by interfacial DMI with scalar spiralization. This picture has been confirmed recently by Koshibae and Nagaosa~\cite{Koshibae_16}, who showed that antiskyrmions are unstable in 2D  chiral  magnets with scalar $D$, but may be realizable as bubbles in magnets with dipolar interaction.

Thus, more general and interesting are crystal symmetries where the spiralization constant $D$ is replaced by a generic tensor quantity $\matt{D}$, which is not included in the $O(2)$ orbit of \eqref{eq:proto}. In this case of anisotropic DMI various scenarios of topological pattern formation including non-symmetric skyrmions and antiskyrmions are possible as shown by means of micromagnetic simulations for combined bulk and interface spin orbit-effects\cite{Rowland_16, Gungordu_16}. 

In the framework of this communication we have extended our analysis \cite{Melcher_14} to two-dimensional chiral systems governed by a micromagnetic energy functional 
\begin{eqnarray}
E=\int_{\R^2} \tfrac{J}{2}|\nabla \m|^2 &&+\matt{D}:\matt{L}(\m) \nonumber \\ 
&&+ B (1- m_z) +K_\perp (m_z^2-1) \, \mathrm{d}\vc{r}.
\end{eqnarray}
with an arbitrary (non-vanishing) spiralization tensor $\matt{D} \in \R^{2 \times 2}$ emerging from \eqref{eq:ham}. Here, 
$\matt{D}:\matt{L}(\m)$
denotes the contraction with the chirality tensor $\matt{L}(\m)=\nabla \m \times \m$ over $x,y$ indices in real and spin space (see Supplementary Note 1 for details). The analysis provides a precise criterion for the optimality of antiskyrmions versus skyrmions
and reveals the diversity of possible chiral phenomena including coexistence of both entities in chiral magnets.\\
The topologically nontrivial magnetization fields are vectorial quantities, assuming values in a multidimensional manifold. Thus, it is conceivable that they are characterized by more than one topological index. 

For skyrmionic configurations with a well-defined skyrmion center $\vc{r}_0$ where $m_z^2 = 1$, it is instructive to take into account the index $v$ of the horizontal magnetization field $m_{\scriptscriptstyle \parallel}=(m_x,m_y)=\sqrt{1-m_z^2} e^{\rm{i} \varphi}$ at $\vc{r}_0$, given by $v=\frac{1}{2\pi} \oint_\Gamma \nabla \varphi \cdot \mathrm{d}\vc{r}$,
which assumes the same integer value for every oriented Jordan curve $\Gamma$ enclosing $\vc{r}_0$. In this case, $v$ can be considered as a secondary topological charge and the defining index to distinguish between skyrmion ($v=1$) and antiskyrmion ($v=-1$) independently of the background state. The index $v$ is also referred to as the $\mathbb{S}^1$ winding number in comparison to the topological charge~\cite{footnote1}, $Q$, which is 
the $\mathbb{S}^2$ winding number. Being independent of $m_z$, the $\mathbb{S}^1$ winding number is independent of the direction of the field polarization $\pm \ein_z$.\par
The DM vector $\vc{D}_{ij}$ imposes a handedness on the magnetic structure 
relating the rotation of the spin $\vc{S}$ about the chirality vector $\vc{c}_\chi(\hat{\vc{R}}_{ij})=\vc{S}_i \times \vc{S}_j$ to the displacement of the spin along the  direction
$\hat{\vc{R}}_{ij} = (\vc{R}_j-\vc{R}_i)/|\vc{R}_j-\vc{R}_i|$ of the bond between the atoms $i$ and $j$. According to our sign convention in \eqref{eq:ham}, the energy is minimized if $\vc{c}_\chi(\hat{\vc{R}}_{ij}) \parallel \vc{D}_{ij}$. We speak of right- (left-) handed Bloch-type skyrmions if $\vc{c}_\chi(\hat{\vc{R}}_{ij})\cdot \hat{\vc{R}}_{ij} >0\, (<0)$, and of N\'eel-type skyrmions if  $\vc{c}_\chi(\hat{\vc{R}}_{ij})\cdot \hat{\vc{R}}^\perp_{ij} >0\, (<0)$,  where $\hat{\vc{R}}^\perp $ is an arbitrary vector of the positive quadrant of the orthogonal complement of $\hat{\vc{R}}$, \eg formed by the vectors $\hat{\vc{e}}_z\times \hat{\vc{R}}$ and $\hat{\vc{e}}_z$.

Although the handedness is not a topological property and thus does not affect topological indices nor the stability of skyrmions, it connects to interesting physical properties. As discussed in more detail below skymions and antiskyrmions are also distinct in their handedness properties.
\\
A double layer Fe on W(110) exhibits a $C_{2v}$ symmetry and is an example with non-zero tensorial components of $\matt{D}$. We show in this communication by means of vector-spin density functional theory (DFT) calculations that its electronic structure leads to long-range and microscopically anisotropic DM vectors that add up to tensorial elements of $\matt{D}$ such that the antiskyrmion becomes the  non-trivial low-energy magnetization. We confirm the stability of antiskyrmions by energy minimization of Eq.~\eqref{eq:ham} on a mesoscopic scale employing atomistic spin-dynamics.
\section{Results}

\begin{figure}
\centering
\includegraphics[width=\linewidth]{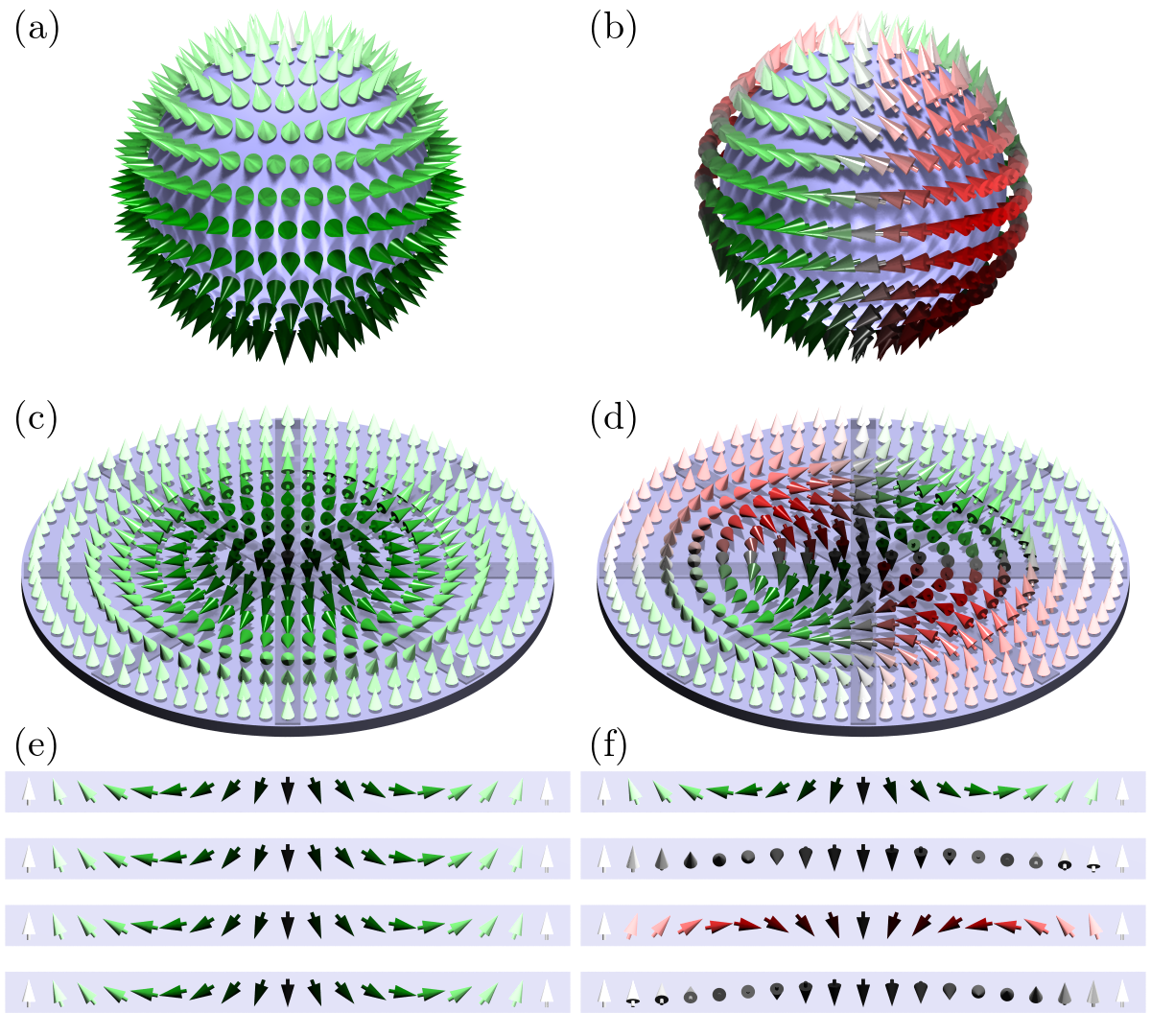}
\caption{\textbf{Comparison of  Skyrmion and  Antiskyrmion.} (a) and (b): N\'eel-like skyrmion and antiskyrmion schematically shown in (c) and (d) mapped onto a sphere. The colour code represents the out-of-plane component of the spins via the brightness, with bright (dark) spins pointing up (down), and their rotational sense in radial direction going from inside-out changing from {red} (clockwise) via {gray} (vanishing rotational sense) to {green} (counter clockwise). (e) and (f): Cross sections of the spin textures along the four highlighted directions shown in (c) and (d), see also Supplementary Movie 1.}
\label{fig1}
\end{figure}

\noindent
\textbf{Symmetry Analysis.}
What may be surprising is that interface stabilized skyrmions have exclusively been explored for (111) oriented interfaces exhibiting a $C_{3v}$ symmetry with the exception of  Mn/W(100), which has a $C_{4v}$ symmetry \cite{Nandy_16}.
Such high-symmetry interfaces may only support N\'eel- or hedgehog-type monochiral skyrmions of positive $\mathbb{S}^1$ winding number, $v=1$, with the magnetization field $\vc{m}(\vc{r})=m_\rho(\rho)\hat{\vc{e}}_\rho+m_z(\rho)\hat{\vc{e}}_z$ when expressed in a cylindrical coordinate system $(\rho, \varphi, z)$ and micromagnetic theory is applied. It is characterized by a fixed sense of cycloidal rotation of magnetic moments from the core outwards independent of the radial direction $\hat{\vc{e}}_\rho$. The sense of rotation can be either clock- or counter-clockwise consistent with the chiral symmetry breaking. It becomes a monopole hedgehog when mapped onto a sphere, see Fig.~\ref{fig1}(a),(c), and (e).

\begin{figure}[t!]
\centering
\includegraphics[width=\linewidth]{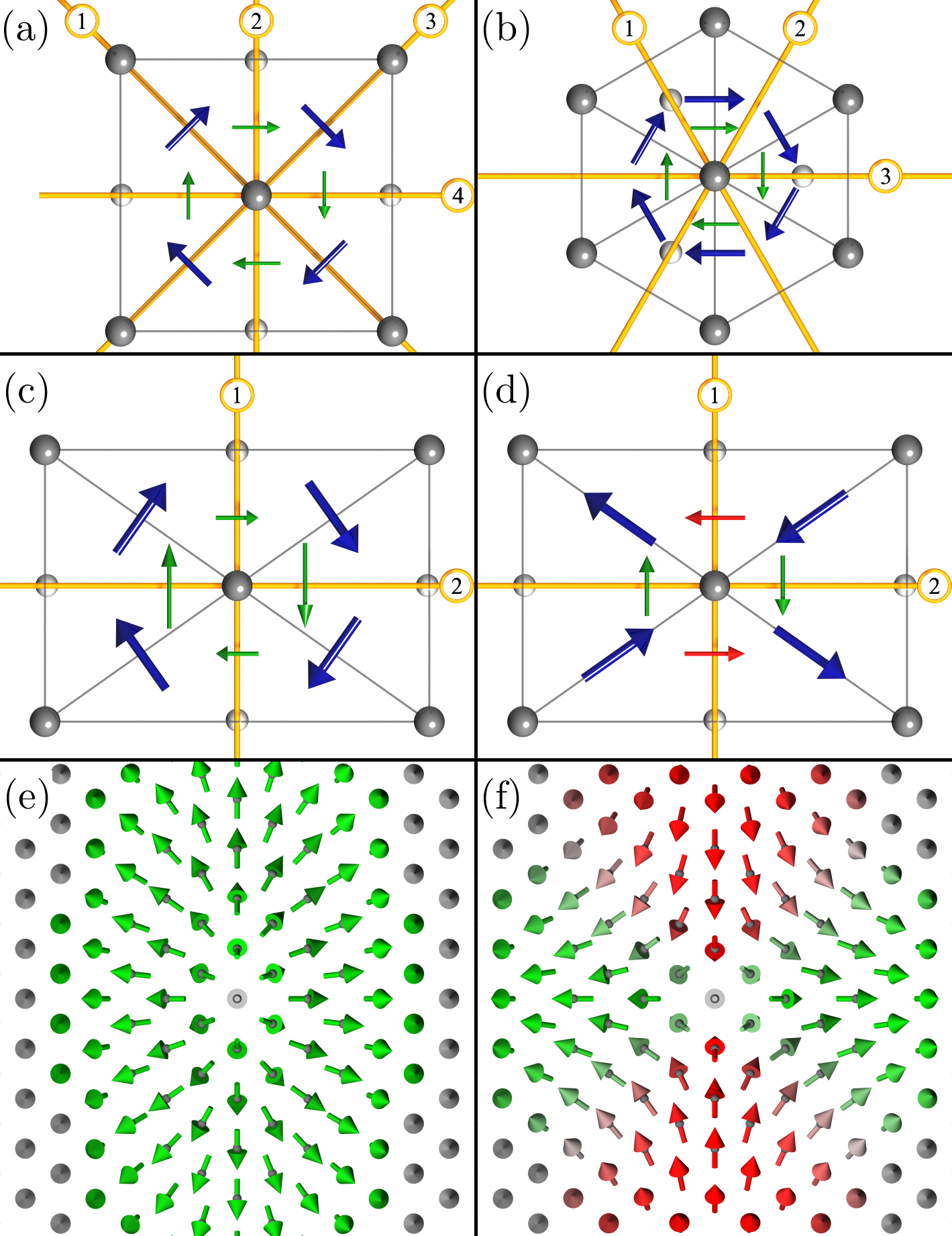}
\caption{\textbf{Visualization of the symmetry to DM vector to magnetization-field relation for different surface unit-cells.} (a) ((b)): Sketch of a square (hexagonal) lattice with $C_{4v}$ ($C_{3v}$) symmetry. Mirror planes are indicated by {orange} lines, atoms by grey balls. The microscopic DM vectors ({blue} arrows) between n.n.\ sites projected onto the crystal surface are in both cases restricted by symmetry to point perpendicular to the bond. The {green} arrows represent the effective micromagnetic DM vector, $\vc{D}_\chi$, enforcing counter-clockwise rotating spin vectors,  $\mathcal{C}_N<0$, in all directions.
(c) and (d): Two particularly chosen examples of DM vectors that match the symmetry rules of Moriya for a system with $C_ {2v}$ symmetry. As discussed in the text and Supplementary Note 4, in principle the two pairs of opposing DMI vectors can point in any in-plane direction.  
(c): as in (a), (b), DM vector to bond relation is the same in all directions. In (d) multichirality occurs, indicated by the two different colours of the effective DM vectors, which result in winding magnetic structures of opposite N\'eel-type chiralities,  $\mathcal{C}_N \gtrless 0$. (e) and (f): Skyrmion and antiskyrmions, with magnetization textures exhibiting the shown preferred rotational senses along the different crystallographic directions. As in Fig.~\ref{fig1}, the colour code indicates the different rotational senses along the radial direction from inside out.
}
\label{fig2}
\end{figure}

Such monochiral skyrmions are formed when the direction of the DM vector relative to the bond direction does not change significantly for a pair of coupled spins in different crystallographic directions.
Examples are systems with $C_{3v}$ or $C_{4v}$ symmetry, which appear typically at (111) or (001) oriented interfaces and  surfaces of bcc and fcc crystals, respectively. This goes all the way back to symmetry arguments for DM vectors, also known as Moriya rules \cite{Moriya}. When applied to these high-symmetry surfaces, they determine uniquely the in-plane components of the direction of the DM vectors, $\hat{\vc{e}}_{\parallel\text{DM}}$, to point perpendicular to the bond connecting the two interacting spins, see in-plane components of the nearest-neighbour (n.n.) DM vectors (blue arrow) $\hat{\vc{e}}_{\parallel\text{DM}} \parallel - \hat{\vc{e}}_{\varphi}$ in Fig.~\ref{fig2}(a) and (b).
Only one scalar parameter $D$ controlling the sign and absolute value along $\hat{\vc{e}}_{\parallel\text{DM}}$ may vary in dependence on the electronic structure.
Furthermore, the mirror symmetries of these surfaces force the DM vectors to have the same $\varphi$-component $D_{\varphi}$ for all symmetry-equivalent atom pairs. Thus, variations of the magnetization along a radial path of any direction $\hat{\vc{e}}_\rho$ will have the same  preferred axis of rotation, $\vc{c}_\chi \parallel -\hat{\vc{e}}_{\varphi}$  (highlighted by the smaller green arrows) when going along different crystallographic directions and therefore only monochiral skyrmions can be formed by the DM interaction.\par
The antiskyrmion (see Fig.~\ref{fig1}(b),(d), and (f)), on the other hand, is the simplest form of a \textit{multichiral} skyrmion \cite{1367-2630-18-5-055015}, which is described by a polar angle $\varphi$-dependent magnetization field $\vc{m}(\vc{r})=m_\rho(\rho,\varphi)\hat{\vc{e}}_\rho+m_\varphi(\rho,\varphi)\hat{\vc{e}}_\varphi+m_z(\rho)\hat{\vc{e}}_z$, with negative  $\mathbb{S}^1$ winding number, $v=-1$,  characterized by different rotational senses along different radial directions away from the core (\cf Fig.~\ref{fig1}(f)), thus showing multiple chiralities. It can be understood naively as an addition of a quadrupolar field to the monopole of the skyrmion (see Supplementary Note 2 and Supplementary Movie 2). Thus, the physical magnetization space has an orientational energy dependence relative to the lattice, a property which may add significance to the role of antiskyrmion applications. 
Due to the multichirality of the magnetic texture, the winding of $m_\rho$ along certain directions $\hat{\vc{e}}_\varphi$ costs energy by the DMI and makes the antiskyrmion seem unfavourable over the skyrmion. 

To form such multichiral structures stabilized by DMI, the DMI evidently needs to show a strong directional dependence.
In films an anisotropic DMI is allowed by symmetries lower than $C_{3v}$. This is schematically sketched in Fig.~\ref{fig2}(c) and (d) for the example of a surface with a centered-rectangular lattice exhibiting $C_{2v}$ symmetry. In this case, the number of symmetry operations is so low that the in-plane direction of the DM vectors, $\hat{\vc{e}}_{\parallel\text{DM}}$, is not anymore uniquely determined by symmetry. Here, the remaining two mirror symmetries  constrain only one component of the DM vector, namely the $D_z$-component to vanish, but the DM vectors are free to point into any in-plane direction.

We recall that distance vectors, $\vc{R}_{ij}$, between pairs of atoms $(i,j)$ form shells of symmetry equivalent vectors of equal length generated by the symmetry operations $\{\matt{R}\}$ mapping atom $j$ onto atom $j'$ and consequently the pseudovector $\vc{D}_{ij'} = \det(\matt{R}) \matt{R} \cdot \vc{D}_{ij}$ (see \eg blue arrows in Fig.~\ref{fig2}(d), which all point along the bond-direction).

\vspace{\baselineskip}\noindent
\textbf{Magnetostatic energy difference between skyrmion and antiskyrmion.} As a consequence of the different magnetisation densities, skyrmions and antiskyrmions embedded in a ferromagnetic background experience a different magnetostatic self-energy $E_\textrm{mag}$.
For an infinite 2D film, $E_\textrm{mag}$ (averaged by the film thickness) decomposes to leading orders into (i) a shape anisotropy energy, $E_\textrm{mag}^\perp$,  and (ii) a magnetic film charge term $E_\textrm{mag}^{\scriptscriptstyle\parallel}$~\cite{Otto_06}. The first term depends solely on $m_z^2$, is independent of the index $v$ and 
accounts for the ferromagnetic dipolar interaction energy included already in the ferromagnetic anisotropy constant $K_\perp$. The second term is proportional to the film thickness $t$, depends on the magnetic film charge $\sigma= (\nabla \cdot m_{\scriptscriptstyle\parallel})$ and is sensitive to the index $v$ and the type of skyrmions characterized by the chirality vector $\vc{c}_\chi(\hat{\vc{R}}_{ij})$. 
$E_\textrm{mag}^{\scriptscriptstyle\parallel}$ as a function of $v$ and $\vc{c}_\chi$ is maximal for N\'eel-type skyrmions, zero for Bloch-type skyrmions and precisely in the middle for axisymmetric antiskyrmions, see Supplementary Note 3.

\vspace{\baselineskip}\noindent
\textbf{Micromagnetic Arguments for skyrmion formation.} The additional degree-of-freedom for the direction $\hat{\vc{e}}_{\parallel\textnormal{DM}}$
may allow for the formation of antiskyrmions over monochiral skyrmions. We will verify this by inspecting in the following the winding of the magnetic structure along $\hat{\vc{e}}_\rho$ from inside out in dependence on $\varphi$ assuming only n.n.\ interaction for two extreme cases of DM vector directions for a system with $C_{2v}$ symmetry: (i) the n.n.\ DM vectors point perpendicular to the bond as shown in Fig.~\ref{fig2}(c) (see blue arrows). In this case the spiralization tensor for 2D systems \cite{PhysRevB.94.024403} takes  the form
\begin{equation}
  \matt{D} = \frac{1}{A_\Omega}  \sum_{j\in(\text{n.n.})}\!\! { \vc{D}_{0j} \otimes \vc{R}_{0j} } = \frac{4 \, \dH}{R}\left( \begin{array}{cc}
   0 & R_y^2 \\
   -R_x^2& 0 
  \end{array}\right),
  \label{eq:spiraliztion_case90}
\end{equation}
where $\vc{R}_{0j} = (R_x,R_y)$ is the distance vector to the n.n.\ atom, $R=\abs{\vc{R}_{0j}}$, $\dH = \abs{\vc{D}_{0j}}$ and $A_\Omega$ is the area per surface atom.

The effective DM vector in the micromagnetic sense,
along the direction of the radial coordinate $\hat{\vc{e}}_\rho = (\cos(\varphi),\sin(\varphi))^\mathrm{T}$
is given by $\vc{D}_\chi=A_\Omega \matt{D}\, \hat{\vc{e}}_\rho$
and indicated in Fig.~\ref{fig2} by thin arrows for four directions of $\hat{\vc{e}}_\rho$  along the four crystallographic $\pm x$- and $\pm y$-axes, respectively. Their green colour indicates that they result all in the same N\'eel-type radial chirality, $\mathcal{C}_N$, which we define as projection of $\vc{D}_\chi$ onto $\hat{\vc{e}}_\varphi$, $\mathcal{C}_N(\varphi) = (\matt{D} \hat{\vc{e}}_\rho)_\varphi$. For the case sketched in Fig.~\ref{fig2}(c) it is always negative, $\mathcal{C}_N = - \frac{4\, \dH}{R} \, \left( R_x^2 \cos^2\varphi + R_y^2 \sin^2\varphi\right)$, irrespective of the radial direction $\hat{\vc{e}}_\rho$, \ie  the green arrows of Fig.~\ref{fig2}(c) show in the $-\hat{\vc{e}}_\varphi$ direction. The independence of $\mathcal{C}_N$ on the radial direction is analogous to interfaces with  $C_{3v}$ and $C_{4v}$ symmetry and stabilizes monochiral skyrmions only (see Fig.~\ref{fig2}(e)).  Since the micromagnetic energy function is minimized if the radial vector chirality of the  spin-structure $\vc{c}_\chi(\hat{\vc{e}}_\rho)$ points parallel to $\vc{D}_\chi$, the negative radial chirality $\mathcal{C}_N$ corresponds to a left- or counter-clockwise magnetic structure along direction $\hat{\vc{e}}_\rho$ as shown in Fig.~\ref{fig2}(e).

(ii) On the contrary, if the n.n.\ DM vectors point parallel to the bonds as in Fig.~\ref{fig2}(d), the off-diagonal entries of \eqref{eq:spiraliztion_case90} change to $-4\dH R_x R_y / R$ and the N\'eel-type chirality  $\mathcal{C}_N(\varphi) = \frac{4\, \dH\, R_x R_y}{R}\left(\sin^2\varphi - \cos^2\varphi\right)$ takes positive \emph{and} negative values, depending on the radial or crystallographic direction of  $\hat{\vc{e}}_\rho$, as indicated by red and green colour, respectively.
It is now evident, that multichiral antiskyrmions are stabilized by such a DM-field.\par
We stress that in general all directions of the DM vector are possible. They are finally system dependent and determined by the details of the electronic structure. Even when the DM vectors deviate considerably from the direction of the bond, multichiral spin-textures remain lower in energy than monochiral ones as we show in the {Supplementary Note 4}. \par

A whole class of systems which possess such a $C_{2v}$ symmetry are magnetic films on the (110) surfaces of bcc or fcc crystals.
Selecting a heavy metal substrate may result in a strong DM interaction \cite{PhysRevB.91.144424}.
In this context the prototype experimental system is certainly an Fe double layer on W(110) \cite{PhysRevLett.87.127201,PhysRevLett.88.057201,PhysRevLett.103.157201,PhysRevLett.104.137203}.
It was shown that the ground state of the system is a ferromagnet with an out-of-plane easy axis, which is favourable for skyrmion stabilization \cite{PhysRevB.90.115427} and it shows a DMI, not strong enough to establish a chiral ground state, but strong enough to stabilize N\'{e}el-type domain walls of unique rotational sense \cite{PhysRevB.78.140403, PhysRevLett.103.157201, PhysRevLett.104.137203}.
Interestingly, a sign change in the micromagnetic DMI along different high-symmetry directions was predicted theoretically for one and two layers of Fe on W(110) \cite{PhysRevB.78.140403}.\par

\vspace{\baselineskip}\noindent
\textbf{Mathematical analysis of (anti-)skyrmion stability and coexistence.}
Relating a low-symmetry interface to a two-dimensional spiralization tensor we could provide arguments that make the stabilization of antiskyrmions plausible. This challenges the arguments of \onlinecite{Melcher_14}, which identify skyrmions as local energy minimizer within the nontrivial topological sector $Q=- 1$, \ie with $\mathbb{S}^1$ winding number $v=1$. 
In the Supplementary Note 1 we generalized the arguments to arbitrary two-dimensional spiralization tensors $\matt{D}$. Central results are the following:
\begin{eqnarray}\label{eq:askyrmion_crit}
\det \matt{D} \,\left\{
\begin{array}{lll} 
< 0 &\text{implies preference of antiskyrmions} \\
> 0 &\text{implies preference of skyrmions,}
\end{array} \right.
\end{eqnarray}
see Theorem 1 and 2 in Supplementary Note 1. In the anisotropic case, where the $\matt{D}$ tensor possesses different singular values $0<D_1<D_2$, skyrmions and antiskyrmions coexist, see Theorem 3 in Supplementary Note 1, with approximately equal energies if $D_1 \approx 0$, \ie $\det \matt{D} \approx 0$, marking a transition from a skyrmionic to an antiskyrmionic phase.

\vspace{\baselineskip}\noindent
\textbf{First-principles calculations.}
So far, most DM parameters calculated from \textit{ab initio} were either limited to n.n.\ interactions \cite{PhysRevLett.115.267210,PhysRevB.92.020401} or to the spiralization \cite{Dupe_14,PhysRevB.78.140403,PhysRevB.90.115427,PhysRevB.94.024403,0953-8984-26-10-104202}, the DM strength in the micromagnetic approach, an approximation where most information about the underlying lattice structure is averaged out. Here, we go beyond these limitations and move to a more realistic description, namely the atomistic spin-lattice model~\eqref{eq:ham}. It includes also the {\em pair-wise} DM interactions beyond the n.n.\ approximation, which we directly calculate by means of density functional theory (DFT) (see Methods) both for atoms within the surface and interface layer but also in-between both representing the intra- and interlayer spin-spin interactions, respectively. Those pair-wise DM vectors together with the exchange constants and the on-site anisotropy are then used in spin-dynamics simulations to investigate the stability of skyrmions and antiskyrmions in the system.\par

Our DFT results for the exchange constants $J_{ij}$ witness a strong ferromagnetic behaviour both within each layer separately (n.n.\ values $J^S_{01}=9.07$~meV and $J^I_{01}=14.24$~meV for the surface and interface layer, respectively) and especially between both layers ($J^{SI}_{01}=42.19$~meV).
Going beyond the n.n.\ approximation allows for a far more accurate description of the physical system (the complete set of values is diagrammed in Supplementary Fig.~4, see Supplementary Note 5). 

\begin{figure}
\centering
\includegraphics[width=\linewidth]{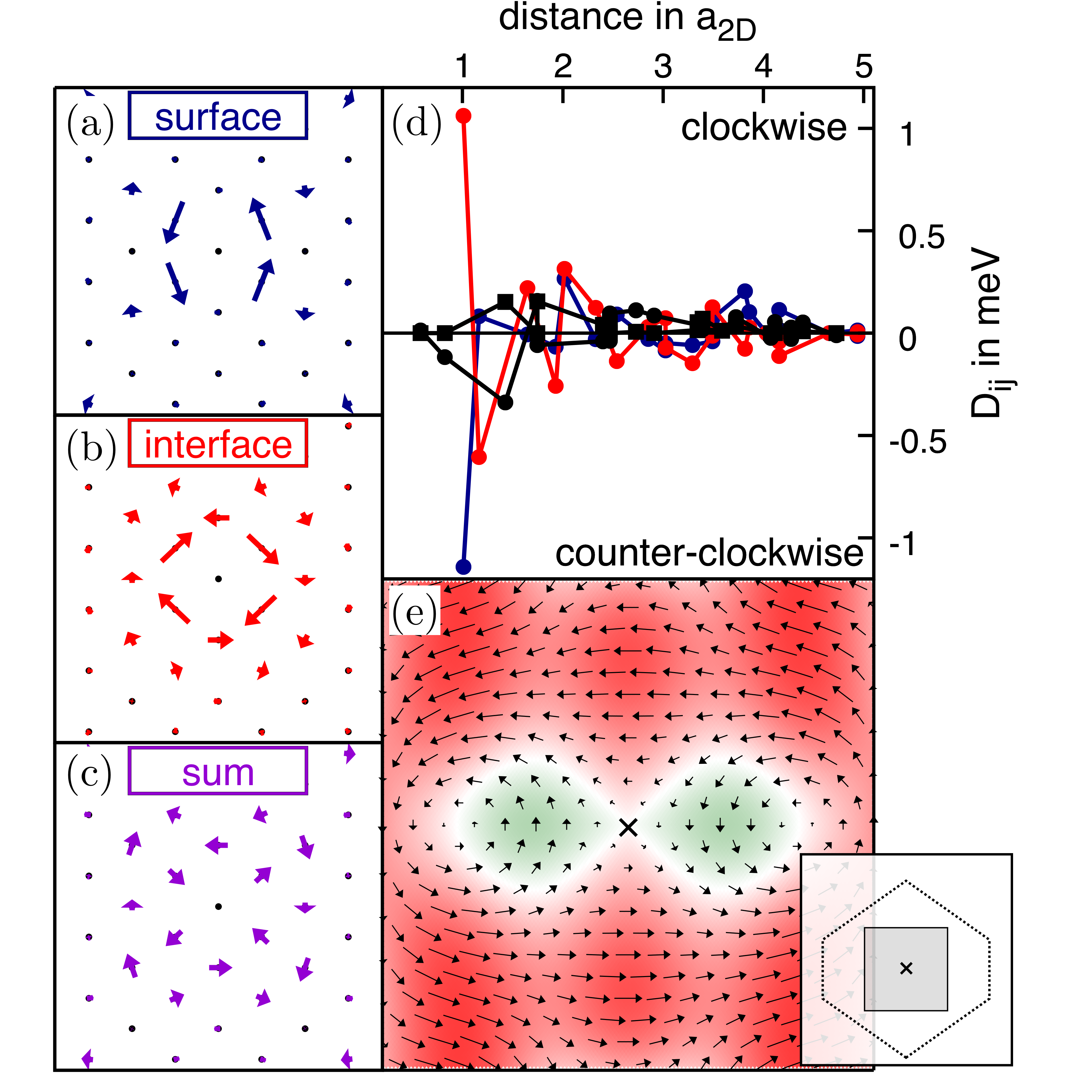}
\caption{\textbf{DM vectors in real and reciprocal space for 2Fe/W(110).} DM vectors $\vc{D}_{0j}$ between atoms at site $0$ and $j$ in (a) the surface, (b) the interface Fe layer and (c) the sum of both superimposed at the surface atoms in 2Fe/W(110), with atom site $0$ at the center of  each panel.  (d):  $\sign{(\vc{D}_{ij})_{\varphi}} \abs{(\vc{D}_{ij})_{\scriptscriptstyle\Vert}}$, with $(\vc{D}_{ij})_{\scriptscriptstyle\Vert}$ being the in-plane component of the DM vector, for the intralayer DM interaction in the surface Fe layer ({blue}), the interface Fe layer ({red}) and the interlayer DM interaction ({black circles}) for the first 21 shells of neighbors. In addition, the out-of-plane component of the interlayer DM interaction is shown ({black squares}). (e): $\vc{D}(\vc{q})$ for $\vc{q}$-vectors close to the center of the two-dimensional Brillouin zone (grey marked area in inset) for a homogeneous spin-spiral in both layers. The colour code represents $(\vc{D}(\vc{q}))_{\varphi}$ going from green (negative values) to red (positive values). DM vectors $\vc{D}_{0j}$ are placed at atom sites $j$ for visual convenience.
} 
\label{fig3} 
\end{figure}

The calculated intralayer DM vectors are depicted in Fig.~\ref{fig3}(a) and (b).
In the surface layer the first n.n.\ DM interaction dominates over all other components, while in the interface layer also the DM vectors of the second n.n.\ are of significant size, but of opposite direction. This oscillatory behaviour can also be seen in Fig.~\ref{fig3}(d), where the distance dependence of 
$\sign{(\vc{D}_{ij})_{\varphi}}\abs{(\vc{D}_{ij})_{\scriptscriptstyle\Vert}}$, \ie with respect to $|\vc{R}_i-\vc{R}_j|$ is diagrammed.  $(\vc{D}_{ij})_{\scriptscriptstyle\Vert}$ is the component of the DM vector parallel to the surface and $\sign{(\vc{D}_{ij})_{\varphi}}$
contributing to the preferred N\'eel-type handedness of the spin texture  ($\sign{(\vc{D}_{ij})_{\varphi}}\rightarrow \sign{\mathcal{C}_N}$. In Fig.~\ref{fig3}(c) the sum of both intralayer DM interactions are superimposed on the surface plane. One can see, that the contributions almost compensate each other (see \eg the first nearest neighbours), and that this compensation leads to a reduction of the absolute value of the effective n.n.\ interaction to the point that the next shells become significant. This shows again the importance to go beyond a n.n.\ approximation to describe this system.

Fig.~\ref{fig3}(e) displays the highly anisotropic nature of the DM vectors of this system expressed in terms of wavevectors in reciprocal space, $\vc{D}(\vc{q})$, and plotted for an area around the $\bar{\Gamma}$-point (ferromagnetic state) of the 2D Brillouin zone (BZ), an appropriate description for films with crystal periodicity.
The slope of $(\vc{D}(\vc{q}))_\varphi$ close to the $\bar{\Gamma}$-point along the different crystallographic directions provides the off-diagonal elements of the spiralization tensor. For 2Fe/W(110) we obtain $\mathcal{D}_{12}= -7.85$~meV/nm ($-1.26$~pJ/m) and $\mathcal{D}_{21}= -6.20$~meV/nm ($-0.99$~pJ/m)~\cite{footnote_3}, respectively, which is in good agreement with the calculated values in \onlinecite{PhysRevB.78.140403}.
The diagonal elements $\mathcal{D}_{11}= \mathcal{D}_{22}=0$ are zero due to $C_\text{2v}$ symmetry and the singular values of $\mathcal{D}$ are $D_1=|\mathcal{D}_{21}|$ and $D_2=|\mathcal{D}_{12}|$, respectively. The determinant $\det\mathcal{D}=-\mathcal{D}_{12}\mathcal{D}_{21}<0$, \ie 
the micromagnetic antiskyrmion condition~\eqref{eq:askyrmion_crit}, which predicts antiskyrmions as the non-trivial topological magnetization soliton of lowest energy, is satisfied. 
In Fig.~\ref{fig3}(e), the different preferred rotation axes, $\vc{c}_\chi(\vc{q})\propto \vc{D}(\vc{q})$, along different crystallographic directions can be clearly seen. However, it also becomes obvious that the preferred rotation axis changes for different periods of spin-spirals even if they propagate along the same direction. With this observation micromagnetic theory might come at its limit. Thus, despite the fact that the micromagnetic criterion for antiskyrmion stability is satisfied, it is advisable to analyse beyond the micromagnetic theory, whether these $\vc{D}$ vectors can lead to the formation of stable antiskyrmions.\par
We turn briefly to the MAE, $K_\perp$ in \eqref{eq:ham}, exhibiting a total out-of-plane easy axis with  $K_\perp=-0.11$~meV/Fe-atom. In addition to the  magnetocrystalline anisotropy ($-0.25$~meV), the dipole-dipole interaction is included here as an effective on-site anisotropy ($0.14$~meV).\par

\vspace{\baselineskip}\noindent
\textbf{Atomistic spin-dynamics simulations.}
To analyse the stability of skyrmions and antiskyrmions in 2Fe/W(110) on a mesoscopic scale we performed atomistic spin-dynamics simulations at zero temperature using the extended first-principles Heisenberg model~\eqref{eq:ham} with parameters determined above (see Methods).

\begin{figure}
\centering
\includegraphics[width=\linewidth]{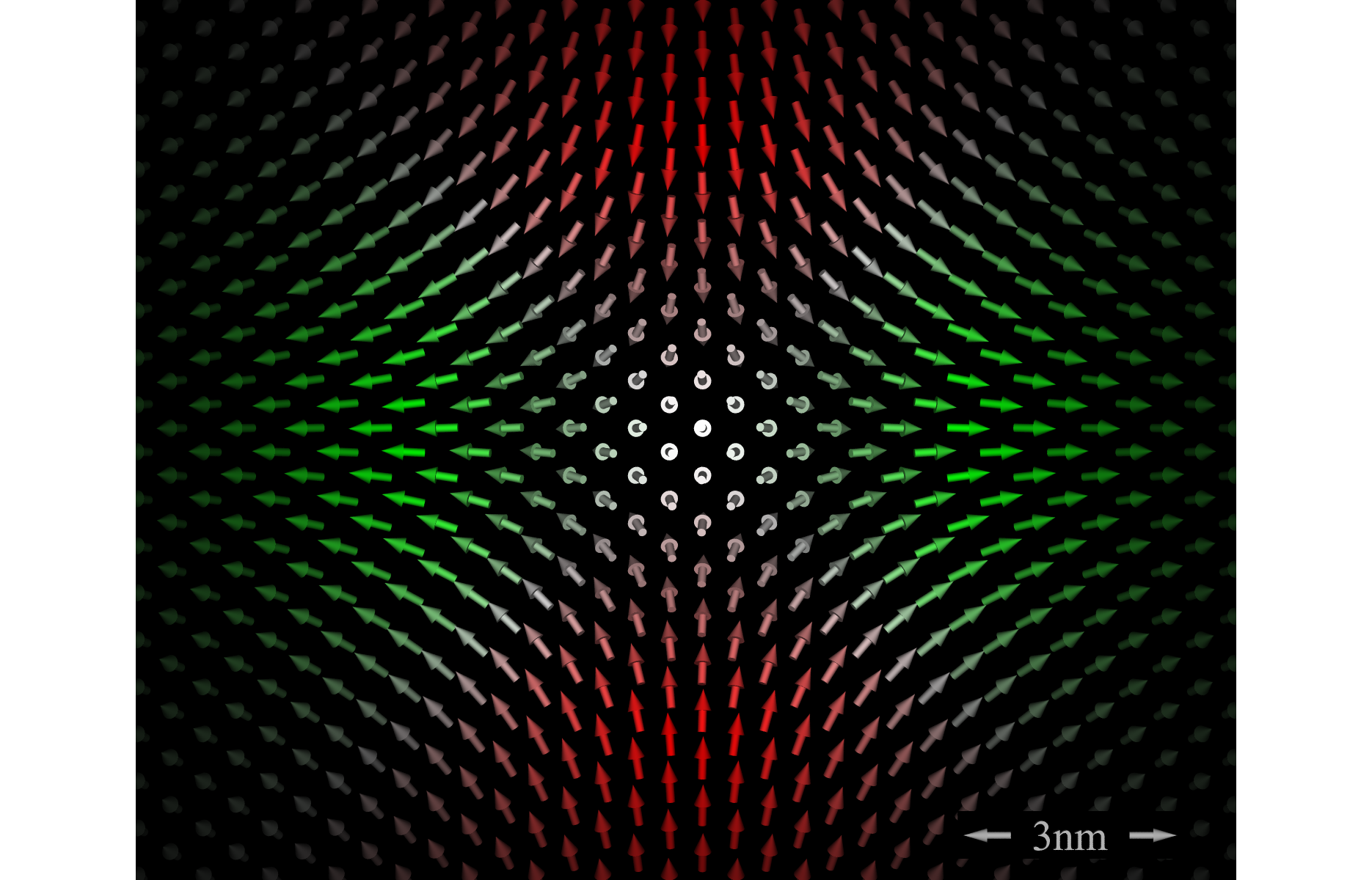}
\caption{\textbf{Top view of a single stable antiskyrmion in a double layer of Fe on W(110) at zero Zeeman field.} While the colour code for the rotational sense of the spins is as in Fig.~\ref{fig1}, the colour code for their out-of-plane component is inverted for visualization purposes. For simplicity, only the Fe surface layer and within this only every second atom are shown. Each arrow represents the direction of the magnetic moment of one single atom. The antiskyrmions resulted from a  simulation including interaction pairs of 8 (9) shells of neighbours for the intralayer (interlayer) coupling equivalent to a total of 26 (26) neighbours.}
\label{fig4}
\end{figure}

While our simulations confirm the ferromagnetic state to be the ground state, antiskyrmions (see Fig.~\ref{fig4} for a top view of the resulting structure) appear as stable states for the interaction parameters describing 2Fe/W(110). They can be found starting from different initial configurations, \eg from antiskyrmions with a broad range of diameters, irrespective of their spatial orientation, as well as from random configurations. We also investigated the stability of the antiskyrmion with respect to the different number of shells of neighbours resolved in the first-principles calculations and subsequently taken into account in the spin-dynamics simulations. Stability of antiskyrmions was found for all tested numbers of shells, however, the exact diameter of the antiskyrmion might vary. Furthermore, for some numbers of shells, a small magnetic field was required to stabilize the antiskyrmion, while for most cases, the antiskyrmion was stable even at zero magnetic field. However, in all tested cases the required field was  lower than $150$~mT, equivalent to about $0.02$~meV/Fe-atom, which lies clearly within the limit of first-principles calculations.

The anisotropy of the micromagnetic $\mathcal{D}$-tensor, $2|{D}_{1}-{D}_{2}|/({D}_{1}+{D}_{2})$ amounts to 23\%. Although the anisotropy is small, our micromagnetic theory suggests a possible coexistence of skyrmions with antiskyrmions.
Starting the atomistic spin-dynamics simulations from any chiral skyrmion,  we observe their collapse into the ferromagnetic state if no external magnetic field is applied. Applying an magnetic field parallel to the core of the skyrmion, we observe in our simulations that at about $700$~mT (the exact value depends again on the number of shells which are taken into account) the skyrmion no longer collapses. Instead, the skyrmion expands towards the boundary of our simulated lattice resulting in a ferromagnetic state aligned parallel to the magnetic field. Fine-tracking the stability of skyrmions with respect to the external magnetic field we can say that within a field resolution of $1$~mT no stable skyrmion was found.
A further analysis of the role of different DM energy contributions (intra- and interlayer coupling) on the stabilization of antiskyrmions can be found in the {Supplementary Note 6}.\par

\section{Discussion}

The interplay of energy and topology resulting in the optimality of certain topological states in micromagnetics is revealed by general principles of mathematical analysis. In the context of 2D systems where DMI is represented by a $2 \times 2$ spiralization tensor $\matt{D}$, it follows that the energetically selected homotopy class (skyrmion versus antiskyrmion) is determined by the orientation of $\matt{D}$, \ie the sign of $\det \matt{D}$ (see Theorem 1 in Supplementary Note 1). Chiral features of the skyrmion texture are determined by the tensorial components of $\matt{D}$. E.g. right- (left-) handed N\'eel-type skyrmions typical for interfaces are obtained if $\sign\left(\mathcal{D}_{21} - \mathcal{D}_{12}\right) = 1\ (-1)$ provided $\mathcal{D}_{21} \ne \mathcal{D}_{12}$.

The prototypical forms of DMI arising in ultra-thin films or from 2D reductions of the cubic B20 type satisfy $\det \matt{D}=D^2>0$ where $D \not=0$ is the DMI constant. As is known \cite{Melcher_14}, chiral skyrmions with $Q=-1$ occur as local energy minimizers within the field polarized regime $BJ > D^2$. The crucial observation is that the least energies $E_Q$ in the topological sectors of charge $Q=\pm 1$ are separated: $E_{-1}< 4 \pi J$ while $E_{1}=4 \pi J$, where the energy quantum $4 \pi J$ is the threshold for the collapse of a topological entity. Consequently, antiskyrmionic configurations may release energy at the expense of forming a point singularity. 

Tuning the system towards $\matt{D}$ with negative determinant, as in the example of an Fe double layer on W(110), lowers the least energy in the topological sector $Q=1$ below $Q=-1$, \ie preference of antiskyrmions. A particularly interesting situation arises in the limit case of vanishing determinant, \ie the presence of effectively only one DM vector, where our analysis predicts the coexistence of skyrmions and antiskyrmions with equal energies. Coexistence, however, is not exclusive to this extreme situation. Whenever the singular values $D_1$ and $D_2$ of $\matt{D}$ differ, the least energies in both topological sectors $Q=\pm 1$ are subcritical $E_{Q}< 4\pi J$. This allows for the occurrence of both entities, while $\det \matt{D}$ serves as a measure of the difference of their energies. 

The magnetostatic energy is not able to change the handedness of the skyrmion, but can be used to tune the energy difference between the skyrmion and antiskyrmion on a fine energy scale as function of the film thickness.

Since the  $\matt{D}$ tensor is a measurable as well as computable quantity, the sign and value of the determinant serves as a simple but powerful criterion for the occurrence and stability of skyrmions and antiskyrmions as well as an effective property descriptor to navigate the search for new materials. It serves also as classification scheme of chiral magnets into isotropic rank-three DMI bulk and rank-two DMI film magnets, with a DMI described by a single constant, the spiralization, for which antiskyrmions are stable only for bulk crystals with certain point group symmetries.  Then, we have the anisotropic rank-two DMI film magnets, where skyrmions and antiskyrmions can coexist, while the sign of $\det \matt{D}$ determines which of the two has the lower energy. Finally, zero determinant indicates a rank-one DMI material, for which skyrmions and antiskyrmions have the same energy.

Our theoretical predictions are consistent with the calculations of G\"ungord\"u \etal\cite{Gungordu_16} observing the occurrence of skyrmion or antiskyrmion lattices depending on details of effective spin-orbit models.  Our analysis now provides a precise selection criterion in the case of general $2 \times 2$ spiralization tensors. Coexistence of isolated skyrmions and antiskyrmions for anisotropic DMI is an entirely new theoretical aspect in the field of chiral skyrmions.

Combining DFT with atomistic  spin-dynamics simulations, in this paper we showed for a double layer Fe on W(110) that the interplay of Heisenberg and anisotropic Dzyaloshinskii-Moriya exchange with uniaxial magnetocrystalline anisotropy and an at most modest external magnetic field results in fact in the stabilization of antiskyrmions. The spatially anisotropic DMI-vectors, $\vc{D}_{ij}$, enabled by the $C_{2v}$ symmetry sum up to a spiralization tensor $\mathcal{D}$, rather than a spiralization scalar $D$, with negative determinant consistent with the micromagnetic antiskyrmion criterion. The tensor is surprisingly isotropic as evidenced by the singular values $D_1\approx D_2$ and no coexistence of skyrmion and antiskyrmion was found by means of  spin dynamics. Details of the electronic structure lead to pair interactions of the DMI of competing sign, showing that a nearest neighbour approximation is insufficient to resolve stability, shape, orientation and fine-structure of the antiskyrmion. As seen from Fig.~\ref{fig4} the symmetry axis of the antiskyrmions matches with the underlying crystal lattice in that the counter-clockwise rotation of the magnetic moments follows the long axis of the unit cell and the clockwise the short one.

Considering that heavy transition-metal substrates~\cite{PhysRevB.90.115427} like W contribute most to the DMI and that their complex Fermi surfaces~\cite{PhysRevB.93.144403} lead to rather anisotropic $\vc{D}_{ij}$ parameters, together with the $C_{2v}$ symmetry we consider 2Fe/W(110) as a prototypical example of a much wider class of magnetic films and heterostructures that host antiskyrmions rather than skyrmions as topologically nontrivial stable states. Other  members of this class are certainly those with (110) oriented interfaces between $3d$ and $5d$ transition-metals. In fact, an anisotropic DMI was recently measured in a thin epitaxial Au/Co film on W(110) \cite{Camosi:arXiv}, but this system favored elliptical skyrmions instead of anti-skyrmions. Other candidates are $3d$ metals on semiconductor (100) and (110) surfaces, \eg Fe on Ge or GaAs, exhibiting  $C_{2v}$ and $C_{s}$ symmetry, respectively.

In this communication we focused on interface stabilized antiskyrmions typical for thin magnetic films and heterostructures, but our arguments apply also to bulk systems with broken inversion symmetry and space groups, for which the effective spiralization tensor is described by a $2\times 2$ matrix rather than a scalar. This motivates the synthesis of new magnetic materials \eg with $C_{2v}$ symmetry on the basis of quaternary selenides such as InVSe$_2$O$_8$\cite{ic200135k}.
When typical chiral bulk magnets, \eg B20 alloys,  with their  lack of bulk inversion symmetry ($\vc{r}\nrightarrow -\vc{r}$) with a DMI described by a spiralization scalar $D$ meet the inversion asymmetry ($z\nrightarrow -z$) of surfaces, films and heterostructures, we expect always a non-zero $\mathcal{D}$-tensor.  Thus, antiskyrmions can emerge at interfaces of skyrmion carrying bulk phases. To complete the discussion, antiskyrmions in films with symmetries resulting in a scalar spiralization do not exist, but can be found in bulk crystals with space groups $D_{2d}$ and $S_4$. The most promising systems to host antiskyrmions seem to be magnetic Heusler alloys (\eg Mn$_2$RhSn~\cite{Meshcheriakova:14}), chalcopyrites (\eg CuFeS$_2$~\cite{Rais2000349}), stannites (\eg Cu$_2$FeSnSe$_4$~\cite{Caneschi2004}) or kesterites (\eg Cu$_2$Mn$_{1-x}$Co$_x$SnS$_4$\cite{LopezVergara2013386}).

Since the topological charge is related to the gyrovector and the Magnus force acting on the antiskyrmion, we expect an antiskyrmion Hall effect. Thus, our work prompts the exploration of the dynamical, dissipative and transport properties related to the additional orientational stabilization of the antiskyrmions with respect to the underlying lattice in comparison to the monochiral skyrmions. 

This work motivates also the design of materials that can host skyrmions and antiskyrmions simultaneously (case $0<D_1<D_2$). This opens an exciting perspective to investigate the interaction of particle and antiparticle of different energies in the spirit of the skyrmion to antiskyrmion interaction in dipolar magnets~\cite{Koshibae_16} or frustrated magnets~\cite{Ezawa_17} and the conditions of mixed ordered lattices and phases.  It should be explored in how far the tunnelling mixing magneto-resistance~\cite{Hanneken:15, Crum:15} (TXMR) effect can be used to discern electrically skyrmions from antiskyrmions. If we optimize materials parameters such that skyrmions as well as antiskyrmions become stable, conditions of mixed ordered lattices and phases may be met. We expect that the topological charge density of an antiskyrmion produces an emergent magnetic field opposite to that of skyrmions for the same material. These different emergent fields give rise to topological orbital moments of opposite sign that can be exploited to discriminate the different skyrmion-antiskyrmion phases spectroscopically using soft X-ray magnetic circular dichroism~\cite{Santos-Dias:2016} (XMCD). These mixed phases are lattices of staggered magnetic fields. The topological Hall effect (THE) of these lattices would be an exciting topic to study.  

Rank-one materials should be particularly interesting for information storage and processing. One may envisage a  creation process of a skyrmion-antiskyrmion pair out of the trivial FM state or a domain wall keeping the total $\mathbb{S}^1$ and $\mathbb{S}^2$ winding numbers, $v$ and $Q$, zero. 
They allow for an extension of the skyrmion race track idea~\cite{Fert_13}, where the information is encoded in the relative positions or time sequences, respectively, of the skyrmions along the track, to the skrymion-antiskyrmion race track memory~\cite{Hoffmann_17}, where the binary information is encoded in the sequence of skyrmions and antiskyrmions, which are expected to be read out distinctly.  Upon further investigations of skyrmion-antiskyrmion interactions in constrictions, rank-one materials may be the ideal host to extend the concepts of the skyrmion logic gates~\cite{Ezawa_15} to magnetic logic gates in which skyrmions and antiskyrmions are the elementary particles for binary operations.

\section{Methods}

\noindent
\textbf{Mathematical analysis.} 
Rather than attempting to solve the equilibrium equations directly, we examine the micromagnetic energy landscape over different homotopy classes. Chiral (anti-)skyrmions arise as relative energy minimizers, which are stable with respect to arbitrary perturbations in the configuration space. The proof of (co-)existence is based on variational arguments combining constructive upper and ansatz-free lower energy bounds of Bogomolny type. The bounds particularly capture the energetic optimality of antiskyrmionic versus skyrmionic configurations (or vice versa) leading to the micromagnetic criterion \eqref{eq:askyrmion_crit}. Crucially in the context of anisotropic DMI, the approach does not rely on any symmetry assumption. Our preliminary symmetry analysis exploits independent $O(2)$ transformations in spin and real space and a singular value decomposition of the spiraliztion tensor (rather than an additive decomposition\cite{Gungordu_16}) to identify a canonical form of DMI. Details are provided in Supplementary Note 1.

\vspace{\baselineskip}\noindent
\textbf{First-principles calculations.}
We performed vector-spin density functional theory (DFT) calculations using the film version of the full-potential linearized augmented plane wave method (FLAPW) as implemented in the \texttt{FLEUR} code~\cite{fleur}. By using the \texttt{FLEUR} code, we get access to the total energy of non-collinear magnetic structures and spin-spirals both with and without spin-orbit coupling, which allows to obtain the parameters entering Eq.~\eqref{eq:ham}. For details on the computations and choice of exchange correlation potential we refer to Supplementary Note 5.

To calculate the pair-wise DM vectors from DFT, we extended the derivation of interatomic exchange interactions in ferromagnets by Le\v{z}ai\'{c} \etal\cite{PhysRevB.88.134403} (which we also used here to calculate the exchange constants $J_{ij}$) to the interatomic DMI parameters $\vc{D}_{ij}$. These have then been evaluated by means of the \texttt{FLEUR} code carrying out coned spin-spirals calculations for a grid of wavevectors $\vc{q}$ with a small cone angle $\theta$ employing the force theorem (for details see Supplementary Note 5).
The results have been cross-checked against independent Korringa-Kohn-Rostocker Green-function (KKR-GF) calculations and we obtained qualitative and quantitative agreement (see Supplementary Note 7).

\vspace{\baselineskip}\noindent
\textbf{Atomistic spin-dynamics simulations.} 
In order to analyse the stability of skyrmions and antiskyrmions we have relaxed approximate and random spin configurations on a lattice with ($150\times 150$) spins in each layer of the Fe double layer at zero temperature according to the Landau–Lifshitz-Gilbert equation of spin dynamics: 
\begin{equation}
\hbar \frac{d\vc{S}_i}{dt}=\frac{\partial H}{\partial \vc{S}_i}\times\vc{S}_i-\alpha\left(\frac{\partial H}{\partial \vc{S}_i}\times\vc{S}_i\right)\times\vc{S}_i
\end{equation}
using the extended Heisenberg model~\eqref{eq:ham}, where $H$ is the Hamiltonian, $\vc{S}_i$ is the spin of unit-length at site $i$  and $\alpha$ is the damping parameter. To achieve fast relaxation from initial distortions of the spin system along the physical path towards lower energy, we have used a range of $\alpha=0.7$ to $0.1$ and time steps ranging from $0.3$ to $3$~fs, respectively. We included the interaction parameters of the neighbours shown in Fig.~\ref{fig3} and in the Supplementary Fig.~4 (see Supplementary Note 5) for the pair-wise contributions including up to 21 shells with in total 71 pairs per atom for both the intralayer and the interlayer coupling. We have carried out simulations for periodic boundary conditions, so as to avoid influence of the boundary on the stabilization of the antiskyrmion.
The equation of motion was integrated by applying an efficient and robust semi-implicit numerical method with built-in angular momentum conservation~\cite{Mentink:10}, as implemented in \texttt{Spirit}~\cite{spirit}. States, which are stabilized this way in spite of the different initial perturbations or distortions, we experienced as local minima in the energy landscape described by the Hamiltonian.

\section{Data availability}
The data that support the findings of this study are available from the corresponding authors upon request.

\section{Acknowledgments}
We would like to thank Miriam Hinzen, Frank Freimuth, Yuriy Mokrousov, and Hannes J\'onsson for fruitful discussions. We gratefully acknowledge computing time on
the JURECA supercomputer provided by the Jülich Supercomputing Centre (JSC).
B.Z.\ and S.B.\ acknowledge funding from the European Union’s Horizon 2020 research and innovation programme under grant agreement number 665095 (FET-Open project MAGicSky).
G.P.M.\ acknowledges funding from the Icelandic Research Fund (grant no. 152483-052). 
Ch.M.\ acknowledges funding from Deutsche Forschungsgemeinschaft (DFG grant no. ME 2273/3-1). 
Ch.M.\ and S.B.\ acknowledge seed-fund support from JARA-FIT.

\section{Author contributions}
M.H.\ and B.Z.\ conceptualized and carried out the DFT calculations. G.M., D.S., M.H., and N.K.\ developed a spin-dynamics code and M.H.\ and G.M.\ performed the spin dynamic simulations. Ch.M.\ provided the mathematical analysis of the micromagnetic functional. S.B.\ initiated this work. All authors took part in the analysis and the discussion of the results and contributed to the writing of the paper.

\section{Competing financial interests}
The authors declare no competing financial interests.


\begin{thebibliography}{99}
\bibitem{footnote_3}
$-3.66 \>$mJ/m$^2$ and $-2.88 \>$mJ/m$^2$ are the respective values in terms of volume quantities, frequently used in micromagnetic simulations. We assume a thickness of $2\cdot d_{Fe-Fe}$ with $d_{Fe-Fe} = 0.172$~nm being the layer distance between the two Fe layers.
\end{thebibliography}
\end{document}


\title{Supplementary Material: Antiskyrmions stabilized at interfaces by anisotropic Dzyaloshinskii-Moriya interaction}

\newcommand{\fz}{Peter Gr\"unberg Institut and Institute for Advanced Simulation, Forschungszentrum J\"ulich and JARA, 52425 J\"ulich, Germany}
\newcommand{\iceland}{Science Institute of the University of Iceland, VR-III, 107 Reykjav\'{i}k, Iceland}
\newcommand{\rwth}{Department of Mathematics I \& JARA FIT,
RWTH Aachen University, 52056 Aachen, Germany}
\author{Markus Hoffmann}
\email{m.hoffmann@fz-juelich.de}
\affiliation{\fz}
\author{Bernd Zimmermann}
\affiliation{\fz}
\author{Gideon P. M\"uller}
\affiliation{\fz}
\affiliation{\iceland}
\author{Daniel Sch\"urhoff}
\affiliation{\fz}
\author{Nikolai S. Kiselev}
\affiliation{\fz}
\author{Christof Melcher}
\affiliation{\rwth}
\author{Stefan Bl\"ugel}
\affiliation{\fz}

\date{\today}

\maketitle

\section{Supplementary Note 1 $\mid$  Mathematical criteria for optimality of antiskyrmions and coexistence with skyrmions}

In this Supplementary Note we rigorously identify simple criteria on the spiralization tensor that predict the energetic optimality of antiskyrmions versus skyrmions (Theorem \ref{thm1}) and the coexistence of skyrmions and antiskyrmions (Theorem \ref{thm3}) under the influence of a sufficiently large Zeeman field, respectively.

\begin{theorem}\label{thm1} Suppose the spiralization tensor has negative determinant. Then for sufficiently large Zeeman field, the least energy over all non-trivial homotopy classes is necessarily attained by an antiskyrmion.
\end{theorem}

The key is to use independent orthogonal transformations in spin space \textit{and} real space in order to bring DMI to a canonical form, which allows to utilize arguments from \onlinecite{Melcher_14} showing that the skyrmion is favoured for the cubic Hamiltonian $D \, \m \cdot \nabla \times \m$. The necessary extension to the 
case of anisotropic DMI
will be provided in Theorem \ref{thm2} below.\\

\textbf{Relative skyrmion number.} The notion of skyrmion/antiskyrmion refers to a local energy minimizer within the topological class characterized by the relative skyrmion number $N=\pm 1$, respectively. For sufficiently regular fields $\m=(m_x,m_y,m_z)$ decaying to the background state $m_z(\infty) =\pm 1$, the index $N$ is defined relative to this background state, \ie
\begin{equation}
N(\m)=-m_z(\infty) Q(\m),
\end{equation}
where $Q$ is the conventional topological charge 
\begin{equation}
Q(\m)= \frac{1}{4\pi} \int_{\R^2} \m \cdot (\del_x \m \times \del_y \m)\, \mathrm{d}\vc{r}.
\end{equation}
In a typical situation where the horizontal magnetization field vanishes at a single point (skyrmion center) the relative skyrmion number $N$ agrees with the index of the horizontal magnetization field at the skyrmion center. It is customary to fix the background state $m_z(\infty)=1$, which leads to the characterization $Q=-1$ for skyrmions and $Q=+1$ for antiskyrmions.\\

\textbf{General form of DMI.} We consider energy densities
\begin{equation}
e_{\rm DM}(\matt{D};\m)=\sum_{\nu} \vc{D}_\nu \cdot (\del_\nu \vc{m} \times \vc{m}),
\end{equation}
where $\vc{D}_\nu$ with $\nu =x,y,z$ are the (micromagnetic) DM vectors, the columns of the 
spiralization tensor $\matt{D}=(\matt{D}_{\mu\nu}) \in \R^{3 \times 3}$, \ie  
$\matt{D}_{\mu \nu}$ is the $\mu$-th component of the $\nu$-th DM vector. For 3D systems, the chirality tensor 
\begin{equation}
\matt{L}(\m)=\nabla \vc{m} \times \vc{m},
\end{equation}
whose components of $\matt{L}(\m)$ are the Lifshitz invariants of $\m$,
can then be used to write DMI in form of a matrix inner product 
\begin{equation}\label{eq:inner}
e_{\rm DM}(\matt{D};\m)= \matt{D}:\matt{L}(\vc{m}) = \sum_{\mu, \nu} \matt{D}_{\mu \nu}\matt{L}_{\mu \nu}(\vc{m}). 
\end{equation}
For 2D systems with $\vc{D}_\nu \perp \ein_z$ for $\nu =x,y$, the spiralization and chirality tensors may be reduced
to $\R^{2 \times 2}$ matrices, respectively. We shall adopt the same notation for the reduced tensors
$\matt{D} \in \R^{2 \times 2}$ and 
$\matt{L}(\vc{m}) \in \R^{2 \times 2}$ given by
\begin{equation}
\matt{L}_{\mu \nu}(\vc{m})=\sum_\kappa \epsilon_{\mu \kappa } \left(m_z \, \del_\nu m_\kappa- m_\kappa \, \del_\nu m_z \right). 
\end{equation}
Here and in what follows greek indices denote in-plane Cartesian coordinates $x,y$, and $\epsilon_{\mu \nu}$ and $\delta_{\mu \nu}$ denote Levi-Civita and
Kronecker symbols, respectively. For spiralization tensors $\matt{D} \in \R^{2 \times 2}$, formula \eqref{eq:inner} features a general micromagnetic form of interface-induced DMI. It also includes the 2D reduction of cubic DMI $\matt{D}_{\mu \nu}= D \, \delta_{\mu \nu}$ differing from the prototypical thin-film DMI $\matt{D}_{\mu \nu}=D \,\epsilon_{\mu \nu}$ only by a rigid rotation in spin space. \\

\textbf{Canonical form of DMI.}
Given orthogonal transformations $\matt{R}, \matt{S} \in O(2)$ in real space and horizontal spin space, respectively, we represent
$
\m(\vc{r})=\matt{S}\tilde{\m}(\tilde{\vc{r}})$ with $\tilde{\vc{r}}=\matt{R}\vc{r}$.
Then, the topological charge satisfies
\begin{equation}\label{eq:QRS}
Q(\m) = \det(\matt{R}\matt{S}) \; Q(\tilde{\m}).
\end{equation}
The chirality tensor satisfies
\begin{equation}
\matt{L(\m)}= (\det \matt{S}) \,\matt{S}  \matt{L}(\tilde{\m})
 \matt{R},
\end{equation}
to be evaluated at $\vc{r}$ and $\tilde{\vc{r}}$, respectively. Defining
the auxiliary spiralization tensor $\tilde{\matt{D}}$ by
\begin{equation}
\matt{D}= (\det\matt{S}) \, \matt{S} \tilde{\matt{D}}\matt{R},
\end{equation}
it follows from \eqref{eq:inner} that
\begin{equation}
\int_{\R^2} e_{\rm DM}(\matt{D};\m) \, \mathrm{d}\vc{r}= 
\int_{\R^2}  e_{\rm DM}(\tilde{\matt{D}};\tilde{\m})\, 
\mathrm{d}\tilde{\vc{r}}.
\end{equation}
Since Heisenberg exchange, uniaxial anisotropy and Zeeman interaction along the vertical direction in spin space are invariant with respect to the transformation between $\m$ and $\tilde{\m}$, the corresponding transformation between $\matt{D}$ and $\tilde{\matt{D}}$ serves as a reduction to a canonical problem. In fact, by virtue of a singular value decomposition we can always achieve $\tilde{\matt{D}}$ to be diagonal and positive semi-definite, leading to the canonical form of DMI
\begin{equation} \label{eq:canonical}
e_{\rm DM}(\tilde{\matt{D}};\tilde \m)= 
D_1 \matt{L}_1(\tilde{\m}) + D_2 \matt{L}_2(\tilde{\m}),
\end{equation}
where $0 \le D_1 \le D_2$ are the singular values of $\matt{D}$ and $\matt{L}_1=\matt{L}_{xx}$ 
and $\matt{L}_2=\matt{L}_{yy}$ the diagonal elements of $\matt{L}$. 
Note that $|\matt{D}|^2=D_1^2+D_2^2$,
where $|\matt{D}|=\sqrt{\matt{D}:\matt{D}}$ denotes the Frobenius norm of $\matt{D}$.\\

In order to prove Theorem \ref{thm1}, we shall argue that in the nondegenerate case $D_1>0$ the transformed problem with DMI \eqref{eq:canonical} selects skyrmions ($Q=-1$). It then follows from \eqref{eq:QRS} that in the original problem with spiraliation tensor $\matt{D}$ antiskyrmions are selected provided $\det (\matt{R} \matt{S})=-1$, \ie if $\det \matt{D}<0$. Theorem \ref{thm1} is therefore a consequence of the following:

\begin{theorem}\label{thm2} Suppose the spiralization tensor is diagonal and positive definite. Then for sufficiently large Zeeman field, the least energy over all non-trivial homotopy classes is necessarily attained by a skyrmion.
\end{theorem}

Let us sketch the proof on Theorem \ref{thm2} concerning DMI as in \eqref{eq:canonical}, \ie $\matt{D}=\mathrm{diag}(D_1,D_2)$. 
We shall assume $D_2 > D_1>0$. The isotropic case $D_1=D_2>0$ is treated in \onlinecite{Melcher_14}.
Dropping the tilde notation we let 
\begin{equation}
e_{\rm DM}(\m)= D_1 \matt{L}_1(\m) +D_2 \matt{L}_2(\m).
\end{equation}
For $B \ge 0$ and $K_\perp \le 0$, we consider the micromagnetic energy functional
\begin{align} \label{eq:micromagnetic}
E(\m)=\int_{\R^2} & \tfrac{J}{2}|\nabla \m|^2 
+ e_{\rm DM}(\m) \nonumber \\
  &+ B (1- m_z) +K_\perp (m_z^2-1) \,
\mathrm{d}\vc{r}.
 \end{align}
 
The key quantities are least energies over homotopy classes characterized by $Q \in \mathbb{Z}$, \ie
\begin{equation}
E_Q= \inf \{E(\m):Q(\m)=Q\},
\end{equation} 
which may not be attained in general.

For $BJ/|\matt{D}|^2 \ge 2$ we claim $E_Q \ge 0$ for all $Q \in \mathbb{Z}$ indicating the ferromagnetic regime. 
Moreover, we have the upper bounds
\begin{equation}\label{eq:upper}
E_{Q}<4\pi J \quad \text{for} \quad Q=\pm 1,
\end{equation} 
approached in the asymptotic limit $BJ/|\matt{D}|^2 \to \infty$.
In particular, $E_{\pm 1}$ lies strictly below the threshold for the formation of a point singularity. In fact, on infinitesimally 
small scales, \eqref{eq:micromagnetic} reduces to Heisenberg exchange, which satisfies a topological lower bound \eqref{eq:classical}
to be used below. In the case $D_1=D_2$ the lower bound 
holds only for $Q=-1$. In addition, we have the lower bounds
\begin{equation} \label{eq:lower}
E_{Q}  > E_{-1} \quad \text{for all} \quad Q \in  \mathbb{Z} \setminus \{0,-1\}.
\end{equation}
In particular, skyrmionic configurations can produce lower energies than any antiskyrmionic configuration.
Notably, in the degenerate case $\det \matt{D}=0$, \ie $D_1=0$, it follows from a reflection argument (see below) that $E_Q=E_{-Q}$ for all $Q \in \mathbb{Z}$.\\

\textbf{Upper energy bounds \eqref{eq:upper}.} For this qualitative estimate, which is only sharp in the large field limit, it is sufficient to take into account equivariant (axisymmetric) configurations. 
Constructing trial profiles for skyrmion cores in the topological class $Q=-1$ with minimal Heisenberg exchange of $4\pi J$ amounts to a modification of the stereographic map, well-establised in the framework of Belavin-Polyakov solitons \cite{Belavin_75}. With $\vc{r}^\perp=(-r_y,r_x)$, we specifically choose 
\begin{equation} \label{eq:core}
\m_{\rm core}(\vc{r})=\left(\frac{2\vc{r}^\perp}{r^2+1}, \frac{r^2-1}{r^2+1} \right),
\end{equation}
featuring a counter-clockwise curling of the horizontal components as preferred by \eqref{eq:canonical}. We have
$\int_{\R^2} \matt{L}_{1}(\m_{\rm core}) \, \mathrm{d}\vc{r} = \int_{\R^2} \matt{L}_{2}(\m_{\rm core}) \, \mathrm{d}\vc{r} =-4\pi$, thus
\begin{equation}
\int_{\R^2} e_{\rm DM}(\m_{\rm core})\, \mathrm{d}\vc{r}=-(D_1+D_2) 4\pi.
\end{equation}
A trial profile of opposite charge
is obtained by reflection in horizontal spin space. For $\overline{\m}=(m_x, -m_y, m_z)$,
we have $Q(\overline{\m})=-Q(\m)$ while $\matt{L}_1(\overline{\m})=-\matt{L}_1(\m)$ and $\matt{L}_2(\overline{\m})=\matt{L}_2(\m)$. We obtain 
\begin{equation}
\int_{\R^2} e_{\rm DM}(\overline{\m}_{\rm core})\, \mathrm{d}\vc{r} 
=(D_1-D_2) 4\pi.
\end{equation}
while the
other energy contributions remain unchanged. Zeeman and anisotropy energies diverge logarithmically for 
\eqref{eq:core}. But using a cut-off and scaling argument to balance all energy contributions (see \onlinecite{Melcher_14}), one deduces \eqref{eq:upper}
provided $D_1<D_2$. The bound holds independently of $B$ but narrows down for increasing $B$. In case of equality $D_1=D_2$ treated in \onlinecite{Melcher_14}, \eqref{eq:upper} follows only for $Q=-1$. \\

\textbf{Ansatz-free lower bounds \eqref{eq:lower}.} 
A straight forward lower energy bound is obtained by bounding the absolute value of DMI in terms of exchange and Zeeman interaction. The argument is independent of the specific form of DMI $e_{\rm DM}(\m;\matt{D}) = \matt{D}:\matt{L}(\m)$ and only depends on its size $|\matt{D}|$. Using the translation invariance of DMI in (vertical) spin space, \ie 
\begin{equation}
\matt{D}:\matt{L}(\m)=\matt{D}:\matt{L}(\m-\ein_z)
+\text{boundary terms,}
\end{equation}
and 
$|\matt{D}:\matt{L}(\m -\ein_z)| \le  |\matt{D}||\nabla \m||\m-\ein_z|$,
we obtain from the elementary inequality $|ab| \le \tfrac{1}{2}(a^2+b^2)$ 
\begin{align}
 \left| \int_{\R^2} e_{\rm DM}(\m;\matt{D}) \, \mathrm{d}\vc{r} \right|
\le  \int_{\R^2} \tfrac{|\matt{D}|^2}{2B} & |\nabla \m|^2 \nonumber \\ + &\tfrac{B}{2}|\m -\ein_z|^2 \, \mathrm{d}\vc{r}.
\end{align}
Taking into account that $ |\m-\ein_z|^2=2(1-m_z)$,  it follows by virtue of the classical topological lower bound  
\begin{equation} \label{eq:classical}
\frac{1}{2} \int_{\R^2} |\nabla \m|^2 \, \mathrm{d}\vc{r}\ge 4\pi |Q(\m)|,
\end{equation}
\cf Belavin and Polyakov\cite{Belavin_75}, that
\begin{equation}
E(\m) \ge 4 \pi  \left(J - \tfrac{|\matt{D}|^2}{B} \right)|Q(\m)|.
\end{equation}
Accordingly, we infer \eqref{eq:lower} for $|Q|>1$ and $BJ/|\matt{D}|^2 \ge 2$.
It remains to show that $E_{-1}<E_{1}$. To this end, we shall modify the Bogomolny type lower bound $E(\m) \ge 4\pi J Q(\m)$ valid for all $\m$ with $Q(\m) \ge 0$ provided $D_1=D_2\not=0$ and $BJ/D_2^2 \ge 1$
(see \onlinecite{Melcher_14}). 
The bound extends to the case $0 \le D_1<D_2$ as
\begin{equation}\label{eq:Bogomolny}
E(\m) \ge 4\pi J Q(\m) -
(D_2-D_1) \int_{\R^2} \matt{L}_1(\m)\, \mathrm{d}\vc{r}.
\end{equation}
We shall argue by contradiction: Suppose $E_1<E_{-1}$ and $\m^{(k)}$ is a minimizing sequence with $Q(\m^{(k)})=1$ and
$E(\m^{(k)}) \to E_1$ as $k \to \infty$. Then it follows from \eqref{eq:upper} with $Q=-1$ and \eqref{eq:Bogomolny} with $Q=1$ that $\int_{\R^2} \matt{L}_1(\m^{(k)})\, \mathrm{d}\vc{r} \ge \lambda$ for some $\lambda >0$ as $k\to \infty$. But for the reflected fields $\overline{\m}^{(k)}$ we have
$
E_{-1} \le E(\overline{\m}^{(k)})  \le E(\m^{(k)}) - 2 D_1 \lambda < E_1
$
as $k \to \infty$, a contradiction. \\

\textbf{Attainment of $E_{\pm 1}$.}
Energy upper and lower bounds can be used to rule out possible scenarios of 
non-attainment of $E_{\pm 1}$ by virtue of the concentration-compactness
method (see e.g. \onlinecite{Lin_04} in the context of the classical Skyrme model).
For $0<D_1=D_2$ we have $E_{Q}<4\pi J$ only for $Q \in \{0,-1\}$, which yields attainment only for $Q=-1$, see \onlinecite{Melcher_14}. Let us now assume $0\le D_1<D_2$. The bounds \eqref{eq:upper} and \eqref{eq:lower} 
then imply 
\begin{eqnarray} \label{eq:splitting}
E_{\pm 1}&<& E_{-Q} + E_{(Q\pm 1)} \quad \text{for all} \quad |Q|>1,\\
\label{eq:concentration}
E_{\pm 1}&<& 4\pi J |Q| +  E_{(Q\pm 1)}  \quad \text{for all} \quad |Q|>0.
\end{eqnarray}
Inequality \eqref{eq:splitting} rules out the splitting into two well-separated field configurations of non-vanishing topological charge 
(dichotomy). Inequality \eqref{eq:concentration} rules out the gain or loss of topological charge by concentration effects. In fact, the formation of one or more point singularities of total charge $Q \not=0$ amounts to an energy of at least $ 4\pi J |Q|$ (see the discussion following \eqref{eq:upper}). Ruling out the vanishing case by means of \eqref{eq:upper} and a Sobolev inequality as in \onlinecite{Melcher_14}, we have exhausted all possible scenarios of non-attainment and obtain:

\begin{theorem}\label{thm3} Suppose the spiralization tensor admits different singular values. Then skymions and antiskyrmions coexist for large enough Zeeman field.
\end{theorem}

\section{Supplementary Note 2 $\mid$ Multipole expansion of an Antiskyrmion}

\begin{figure}
\centering
\includegraphics[width=0.9\linewidth]{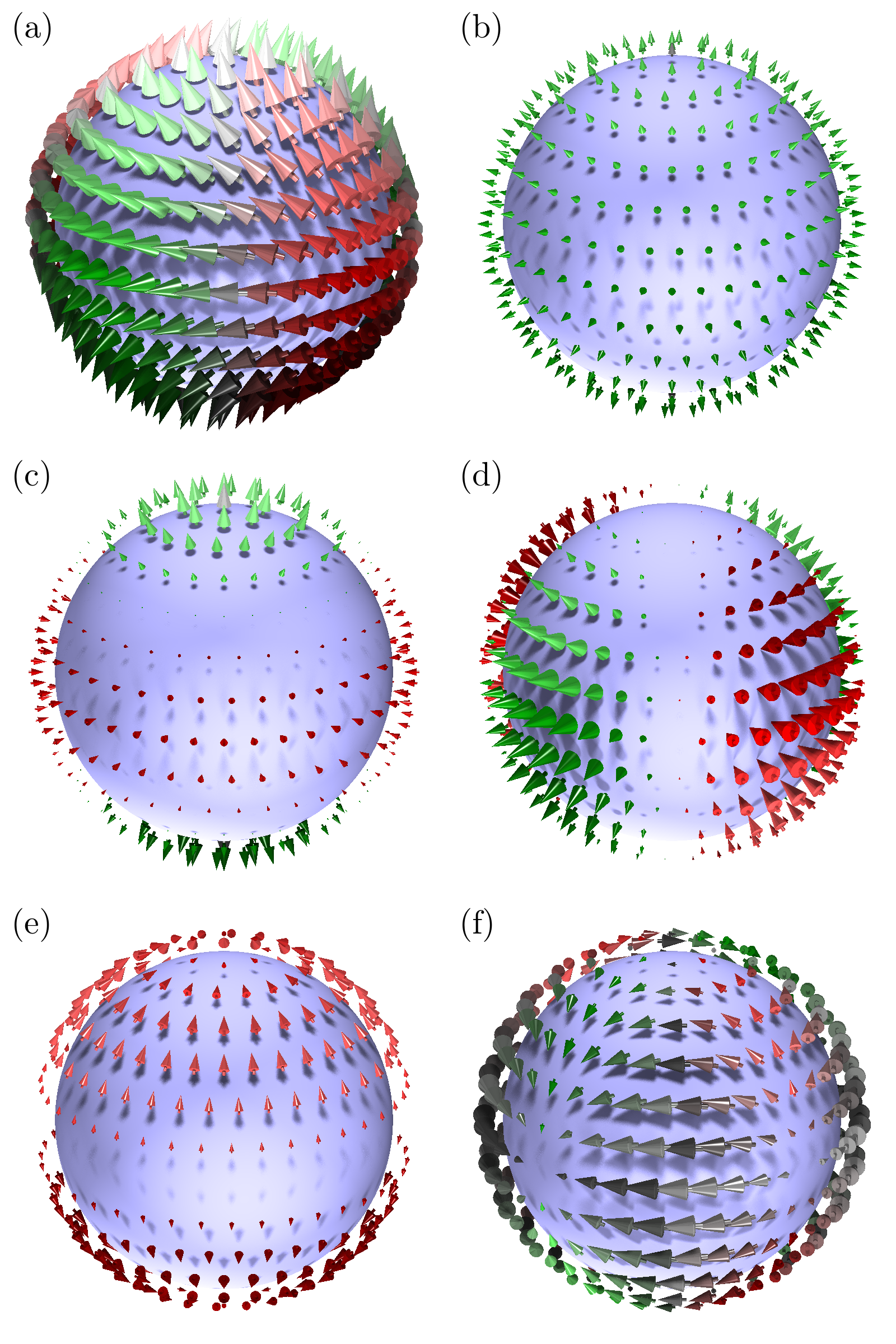}
\caption{\textbf{Visualization of the multipole expansion of an antiskyrmion.} Non-zero contributions of the multipole expansion of the antiskyrmion shown in (a): (b) $\vc{Y}_{00}$, (c) $\vc{Y}_{20}$, (d) $\vc{Y}_{2-2} + \vc{Y}_{22}$, (e) $\vc{\Psi}_{20}$, and (f) $\vc{\Psi}_{2-2} + \vc{\Psi}_{22}$. The length of the arrows indicates the size of the expansion coefficients (see text).}
\label{fig:multipole}
\end{figure}

In the main text we stated that an antiskyrmion can be understood as an addition of a quadrupolar field to the monopole of the skyrmion. This statement will be substantiated in this Supplementary Note. We will show in the following the multipole expansion of the magnetization field of the antiskyrmion on a sphere (see Supplementary Fig.~\ref{fig:multipole}(a)) in terms of the vector spherical harmonics which are defined as\cite{0143-0807-6-4-014}
\begin{eqnarray}
\vc{Y}_{lm} &=& Y_{lm}\, \hat{\vc{e}}_{\vc{r}} \nonumber\\
\vc{\Psi}_{lm} &=& r \nabla Y_{lm}\nonumber\\
\vc{\Phi}_{lm} &=& \hat{\vc{e}}_{\vc{r}} \times \nabla Y_{lm}\, .
\end{eqnarray}
It is obvious, that a hedgehog-type chiral skyrmion as shown in Fig.~1(a) of the main text has the monopole ($\vc{Y}_{00}$) as the only non-vanishing component with an expansion coefficient $c_{00}^{\vc{Y}} = M \sqrt{4\pi}$, with $M = \abs{\vc{m}(\vc{R})}$, being the absolute value of the magnetization. For the antiskyrmion, however, this monopole contribution is reduced ($c_{00}^{\vc{Y}} = \frac{1}{3}\ M \sqrt{4\pi}$) and additionally quadrupolar contributions arise. The resulting expansion coefficients are $c_{20}^{\vc{Y}} = \frac{\sqrt{20}}{15}\ M\sqrt{4\pi}$, $c_{2-2}^{\vc{Y}} = c_{22}^{\vc{Y}} = -\frac{\sqrt{30}}{15}\ M\sqrt{4\pi}$, $c_{20}^{\vc{\Psi}} = \frac{\sqrt{5}}{15}\ M\sqrt{4\pi}$, and $c_{2-2}^{\vc{\Psi}} = c_{22}^{\vc{\Psi}} = - \frac{\sqrt{30}}{30}\ M\sqrt{4\pi}$. The non-vanishing multipole components are visualized in Supplementary Fig.~\ref{fig:multipole}, where the length of the arrows scales with the expansion coefficient.

\FloatBarrier

\section{Supplementary Note 3 $\mid$ Magnetostatic energy of axisymmetric skyrmions vs. antiskyrmions.}

In non-dimensionalized form, the averaged magnetostatic energy $E_{\rm mag}(\m)$ induced by a magnetization distribution $\m$ on an infinite film of thickness
$t$ is given by
\begin{equation}
 \frac{1}{8 \pi t}  \int_{\R^3} \int_{\R^3}  \frac{(\nabla \cdot \m)(\vc{r}) (\nabla \cdot \m)(\vc{r}')}{|\vc{r}-\vc{r}'|} \mathrm d\vc{r} \mathrm d\vc{r}',
\end{equation}
where $\nabla \cdot \m$ is the distributional divergence of $\m$ extended by zero outside the film.
If $\m$ is assumed to be $z$-independent inside the film, the contributions to $E_{\mathrm{mag}}(\m)$ from $m_z$ and $m_{\scriptscriptstyle\parallel}$ separate.
To leading orders in $t$, this gives rise to a decomposition $E_{\mathrm{mag}}(\m) \approx E_{\mathrm{mag}}^\perp(m_z)+E_\mathrm{mag}^{\scriptscriptstyle\parallel}(\sigma)$
into a shape anisotropy part
\begin{equation}
E_{\mathrm{mag}}^\perp(m_z)= \frac{1}{2} \int_{\R^2} m_z^2 \mathrm d\vc{r} 
\end{equation}
and a film charge part
\begin{equation}
E_\mathrm{mag}^{\scriptscriptstyle\parallel}(\sigma)= \frac{t}{8 \pi} \int_{\R^2} \int_{\R^2} \frac{\sigma(\vc{r}) \sigma( \vc{r}')}{|\vc{r}-\vc{r}'|} \mathrm d\vc{r} \mathrm d\vc{r}' 
\end{equation}
in terms of the in-plane divergence $\sigma(\vc{r})= (\nabla \cdot m_\parallel)(\vc{r})$. 
This thin film reduction~\cite{Otto_06} can conveniently be derived from a Fourier space representation, in which the film charge part admits the form \begin{equation} \label{eq:Fourier_charge}
E_\mathrm{mag}^{\scriptscriptstyle\parallel}(\sigma) = \frac t 4 \int_{\R^2} \frac{|\hat{\sigma}(\vc{k})|^2}{|\vc{k}|}  \mathrm{d}\vc{k}.
\end{equation}
We shall estimate the contribution of skyrmions and antiskrymions to $E_{\rm mag}(\m)$ considering the standard axisymmetric ansatz 
with magnetization densities of the form~\cite{Bogdanov:94, Nagaosa:13}
\begin{equation}\label{eq:axisymmetric}
\vc{m}(\vc{r}) = \begin{pmatrix}
\cos\Phi(\varphi)\sin\theta(\rho) \\ \sin\Phi(\varphi)\sin\theta(\rho) \\ \cos\theta(\rho)
\end{pmatrix}
\end{equation}
with cylindrical coordinates $(\rho,\varphi)$ in real and polar coordinates $(\theta, \Phi)$ in spin space, depending on 
$\rho$ and $\varphi$, respectively. The phase function has the form
\begin{equation}
\Phi(\varphi) = v\varphi +\gamma
\end{equation}
with winding number $v=\pm 1$ to distinguish skyrmions from antiskyrmions, and a phase shift $\gamma$ to tune chirality  according to the chirality vector $\vc{c}_\chi$.
With the left-handed N\'eel-type skyrmion $\m_{\text{N\'eel}}$ (with $v=1$ and $\gamma=0$) as reference configuration,
an arbitrary $\m$ of the form \eqref{eq:axisymmetric} is obtained by an orthogonal transformation in horizontal spin space, \ie
\begin{equation}
\m = \matt{S} \m_{\text{N\'eel}} \quad \text{for some} \quad \matt{S} \in O(2),
\end{equation}
keeping shape anisotropy invariant. 
With in-plane divergences of $\m$ and $\m_{\text{N\'eel}}$ denoted by $\sigma$ and $\sigma_{\text{N\'eel}}$, respectively, it follows from a
symmetry argument in \eqref{eq:Fourier_charge} that
\begin{equation} \label{eq:mean}
\frac{E_\mathrm{mag}^{\scriptscriptstyle\parallel}(\sigma)}{E_\mathrm{mag}^{\scriptscriptstyle\parallel}(\sigma_{\text{N\'eel}})} =  \frac{1}{2\pi} \int_{\{|\vc{k}|=1\}} |\vc{k} \cdot (\matt{S} \vc{k})|^2 \mathrm{d}\vc{k}.
\end{equation}
Expressed in terms of $v$ and $\gamma$ \eqref{eq:mean} reads
\begin{equation}
\frac{E_\mathrm{mag}^{\scriptscriptstyle\parallel}(\sigma)}{E_\mathrm{mag}^{\scriptscriptstyle\parallel}(\sigma_{\text{N\'eel}})} = \begin{cases} \cos^2\!\gamma \;\text{for skyrmions} \; v=1,\\
 \frac 1 2 \;\text{for antiskyrmions} \; v=-1. \end{cases}
\end{equation}
Accordingly, the effect of film charge differs significantly for different kinds of chiral skyrmions. It is minimal (zero) for Bloch-type ($\gamma=\pm \frac \pi 2$), and maximal for N\'eel-type skyrmions ($\gamma=0,\pi$).
Axisymmetric antiskyrmions are precisely in the middle with a value independent of $\gamma$. The argument may be extended to the case of anisotropic DMI and almost axisymmetric (anti-)skyrmions near the conformal high energy limit~\cite{Doering_Melcher:17} $JB/|\matt{D}|^2 \gg 1$.

\section{Supplementary Note 4 $\mid$ Multichirality in systems with the $C_{2v}$ symmetry from a micromagnetic analysis}

\begin{figure}[t]
\centering
\includegraphics[width=\linewidth]{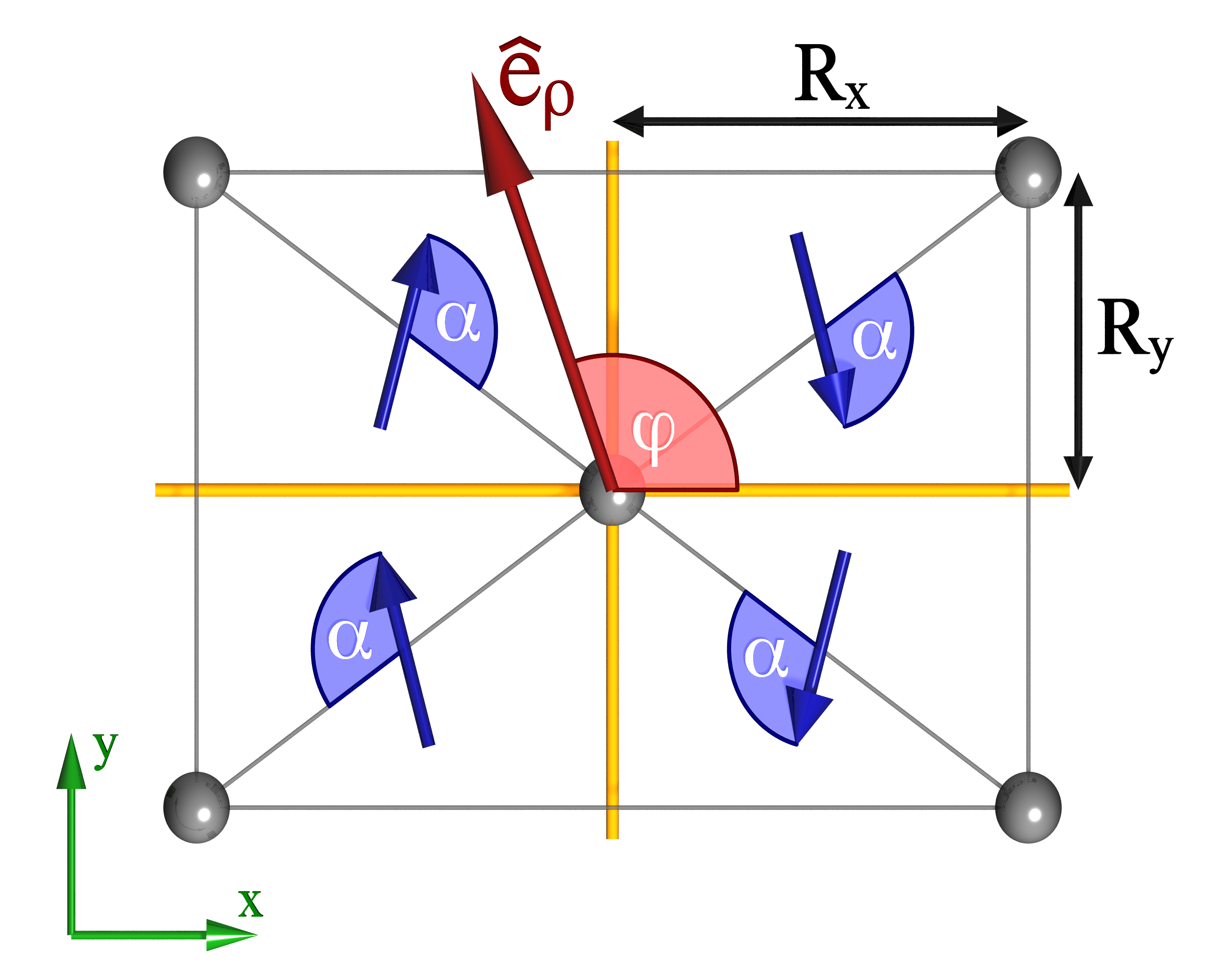}
\caption{\textbf{Visualization of the investigated geometry.} Visualization of the angles and distances introduced in the text for the description of a system with $C_{2v}$ symmetry.}
\label{fig:appendix-angles}
\end{figure}

In the main paper we have shown that for surfaces or interfaces with $C_{2v}$ symmetry, the symmetry is so low that the in-plane component of the DMI vector, $\hat{\vc{e}}_{\parallel\text{DM}}$, is not symmetry-determined, but can take on any in-plane direction depending on details of the electronic structure of the system. We had selected two particular directions of $\hat{\vc{e}}_{\parallel\text{DM}}$, one resulting in a skyrmion and one in an antiskyrmion. In this supplementary note, we generalize the equations to arbitrary in-plane directions and to any system with a $C_{2v}$ symmetry.

In a system with $C_{2v}$ symmetry, the general form of the spiralization tensor reads\cite{Masterarbeit_hinzen}
\begin{equation}
  \matt{D} =  \left( \begin{array}{cc}
   0 &  \mathcal{D}_{12} \\
   \mathcal{D}_{21} & 0
  \end{array}\right)\, ,
  \label{eq:Dtensor-C2v}
\end{equation}
where $\mathcal{D}_{12}$ and $\mathcal{D}_{21}$ contain all information about the geometry of the lattice and result from the summation over all neighbours. Note, that $\det\mathcal{D}=-\mathcal{D}_{12}\mathcal{D}_{21}$ and $\mathcal{D}_{\mu\nu}$, $\mu\ne \nu$, can be related to the singular values of the Supplementary Note 1 by $D_{1(2)} = \min(\max)(\abs{\mathcal{D}_{12}},\abs{\mathcal{D}_{21}})$. It is insightful to map the spiralization tensor~\eqref{eq:Dtensor-C2v} onto a nearest-neighbour model (see Supplementary Fig.~\ref{fig:appendix-angles}) with an effective n.n.\ DM vector, $\vc{D}_\text{nn}$. The two linearly independent components of $\vc{D}_\text{nn} = (D_x, D_y)$, translate into two independent parameters of the spiralization tensor, which becomes
\begin{equation}
  \matt{D} =  \frac{1}{A_\Omega}\left( \begin{array}{cc}
   0 & 4 D_x R_y \\
   4 D_y R_x & 0
  \end{array}\right).
  \label{eq:Dtensor}
\end{equation}
This spiralization tensor is obviously equivalent to \eqref{eq:Dtensor-C2v} with $\mathcal{D}_{12} = {4 D_x R_y}/{A_\Omega}$ and $\mathcal{D}_{21} = {4 D_y R_x}/{A_\Omega}$ and is therefore capable to describe the identical mathematical problem.
We express the direction of the microscopic DM vector,
\begin{equation}
  \vc{D}_\text{nn} =  \left( \begin{array}{c}
   D_x \\
   D_y
  \end{array}\right) = \frac{\dH}{R} \left( \begin{array}{r}
   \cos \alpha ~ R_x + \sin \alpha ~ R_y \\
  -\sin \alpha ~ R_x + \cos \alpha ~ R_y
  \end{array}\right),
\end{equation}  
relative to the vector of the chemical bond connecting the atom at site $\vc{R}_0$ and the n.n.\ sites $\vc{R}_\text{nn}$, $\vc{R}_\text{nn}=(R_x, R_y)$, in terms of angle $\alpha$, 
where $\dH = \sqrt{D_x^2 + D_y^2}$ and $R = \sqrt{R_x^2 + R_y^2}$ and $1/A_\Omega$ is the area of the surface unit cell. Hence, the angles $\alpha = \pi/2$ and $\pi$ reproduce the spiralization tensors discussed in the main text.
The introduced angles and the lattice parameters are visualized in Supplementary Fig.~\ref{fig:appendix-angles}.

According to the main text, the N\'eel-type chirality, $\mathcal{C}_N$, for a magnetic winding along direction  $\hat{\vc{e}}_\rho = (\cos \varphi, \sin \varphi )^\mathrm{T}$ is then given by 
\begin{eqnarray}
\hspace{-1cm}  \mathcal{C}_N &=& A_\Omega (\matt{D} \hat{\vc{e}}_\rho)_{\varphi} \nonumber \\
    &=& \phantom{-} \frac{4\,\dH}{R}R_x R_y \cos\alpha ~ \cos(2\varphi)  \nonumber \\
    & & - \frac{4\,\dH}{R}R_x R_y\sin\alpha \left( \cos^2\!\varphi ~ \frac{R_x}{R_y}  + \sin^2\!\varphi ~ \frac{R_y}{R_x}  \right) \label{eq:app:chirality}
\end{eqnarray}
For a fixed angle $\alpha$, only the first term $\propto \cos2\varphi$ can change the sign of the chirality $\mathcal{C}_N(\varphi,\alpha)$ as function of $\varphi$, while the second term supports monochirality. Hence, $\alpha$ weights the relative competition between multi- and monochirality, or antiskyrmion and skyrmion formation, respectively.

We now restrict ourselves to the geometry of a bcc(110) surface, for which $R_x=a/\sqrt{2}$ and $R_y=a/2$, with $a$ being the lattice parameter. Analysing $\mathcal{C}_N(\varphi,\alpha)$ as a function of the direction of the in-plane DM vector in terms of $\alpha$ and the direction of $\hat{\vc{e}}_\rho$ parametrized by $\varphi$, we see in Supplementary Fig.~\ref{fig_chirality_bcc110} that the DM-field changes sign as function of $\varphi$ for a broad range of $\alpha$, \ie $-\arctan(\sqrt{2}) < \alpha < \arctan(1/\sqrt{2})$, thus favouring magnetization fields of multichiral character. Outside this parameter range of $\alpha$ we have monochiral DM-fields either of positive ($\alpha >\arctan(1/\sqrt{2}$,  $\mathcal{C}_N>0$) or negative ($\alpha >\arctan(1/\sqrt{2}$,  $\mathcal{C}_N>0$) chirality.
\begin{figure}
\centering
\includegraphics[width=\linewidth]{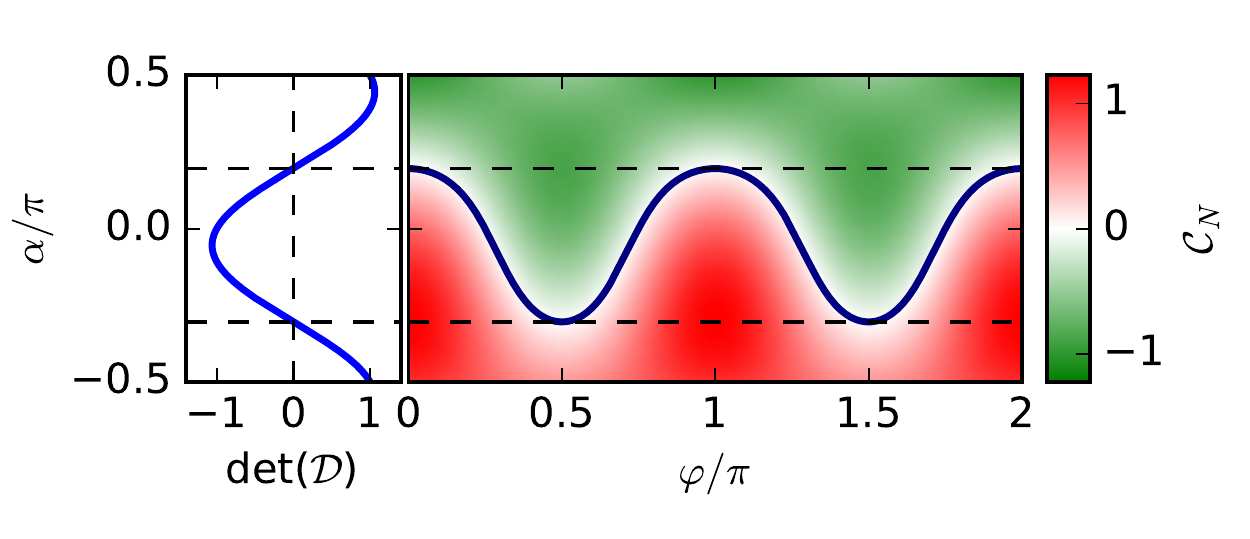}
\caption{\textbf{N\'eel-type chirality based on the micromagnetic model.} N\'eel-type chirality (as colour code) for a bcc(110) surface as function of the relative angle $\alpha$ between the DM vector and the bond between atoms, and of $\varphi$ parametrizing  the propagation directions $\hat{\vc{e}}_\rho$. Regions of positive and negative chirality are separated by a solid line. For a broad range of $-0.3\pi \lesssim \alpha \lesssim 0.2 \pi$ (indicated by horizontal dashed lines) multichiral skyrmion states are possible. In this range, the determinant of the spiralization tensor, $\det{\mathcal{D}}$, is negative. Values for $\mathcal{C}_N$ have been scaled by a factor $\sqrt{3}/(a D)$ and $\det{\mathcal{D}}$ by $A_\Omega^2 4\sqrt{3}/(a^3 D)$.}
\label{fig_chirality_bcc110}
\end{figure}

This analysis is consistent with the criteria given in Supplementary Note 1.
The determinant of the spiralization tensor \eqref{eq:Dtensor} results in
\begin{eqnarray}
\det{\matt{D}} &=& -\frac{16D^2}{R^2 A_\Omega^2}  R_x^2 \, R_y^2 \, \cos(2 \alpha) \nonumber\\ 
& & -\frac{8D^2}{R^2 A_\Omega^2} R_x \, R_y (R_y^2 - R_x^2)  \, \sin(2\alpha)\, .
\end{eqnarray}
The determinant is equal to zero, $\det{\matt{D}}=0$, for angles
\begin{equation}
\alpha_o = \arctan\left(\frac{R_y}{R_x}\right) + n \cdot \frac{\pi}{2}\quad \mathrm{with}\quad n \in \mathbb{Z}\, .
\label{eq:alpha0}
\end{equation}
This reproduces the previous results $\alpha_o = -\arctan(\sqrt{2})$ and $\alpha_o = \arctan(1/\sqrt{2})$ for values of  $R_x$ and $R_y$ given by the bcc(110) lattice. 
In between those two values, $\det{\matt{D}}$ becomes \textit{negative} (see Supplementary Fig.~\ref{fig_chirality_bcc110}) preferring consistent with the conditions derived in the Supplementary Note 1 the antiskyrmion over the skyrmion.
Taking into account our definition of the angle $\alpha$, $\alpha_o$ denotes DM vectors pointing parallel to one of the mirror planes of the system, \ie parallel to the x- or y-axis of the coordinate system shown in Supplementary Fig.~\ref{fig:appendix-angles}.
This can easily be seen for the special case of $R_x = R_y$ (\ie a square lattice) where one obtains $\alpha_o = \frac{\pi}{4} + n \cdot \frac{\pi}{2}$. However, one should keep in mind, if $R_x = R_y$ then usually this change is related to a structural transition from $C_{2v}$ to $C_{4v}$ symmetry where additional mirror symmetries (see arguments in main text) only allow $\alpha_{C_{4v}} = \frac{\pi}{2} + n \pi$. Thus, the case $\det{\matt{D}} = 0$ is forbidden and furthermore it follows  $\det{\matt{D}} = -4 D^2 R^2 \cos(2 \alpha_o)/A_\Omega^2 = 4 D^2 R^2/A_\Omega^2 > 0$ leading only to skyrmions.

\section{Supplementary Note 5 $\mid$ Computational details of DFT calculations for 2Fe/W(110)}

We performed DFT calculations employing the non-collinear version~\cite{Kurz:04} of the full-potential linearized augmented plane-wave method \cite{Wimmer:81.1, Weinert:82.1} (FLAPW) in film geometry as implemented in the \texttt{FLEUR} code \cite{fleur_supp}. The structural properties such as the lattice constant and the interlayer relaxation were taken from Heide \textit{et al.}~\cite{PhysRevB.78.140403}. An asymmetric slab consisting of seven W layers and two Fe layers was used as structural model. A self-consistent calculation without spin-orbit coupling was carried out for the ferromagnetic state with 1600 k-points in the full two-dimensional (2D) Brillouin zone (BZ) and a plane-wave cutoff of 4.2 bohr$^{-1}$ serving as starting point for the non-collinear calculations to follow. An exchange correlation potential in local density approximation \cite{MJW} was used. 

Subsequently, calculations of the electronic structure and total energy were performed for spin-spiral states for a set of wave vectors $\vc{q}$. Thus, for a spin-spiral the magnetic moment vector, $\vc{M}$, at site $\vc{R}$ is determined by
\begin{align}
  \vc{M}(\vc{q}) &= M ~  \matt{R}(\hat{\vc{n}}) ~ \vc{S}(\vc{q})\nonumber\\
   &= M ~ \matt{R}(\hat{\vc{n}})~ \left( \begin{array}{c} \sin\theta ~ \cos(\vc{R}\cdot\vc{q} + \phi) \\
\sin\theta ~ \sin(\vc{R}\cdot\vc{q} + \phi) \\
\cos\theta
\end{array} \right) \, ,\label{eq:ansatz:magneticmoment}
\end{align}
where $\matt{R}(\hat{\vc{n}})$ is a rotation matrix that rotates the rotation axis $z$ of the above equation into the actual rotation axis $\hat{\vc{n}}$ around which the spin-spiral rotates. $\phi$ enables phase differences in the spirals between the first and second layer. The cone angle $\theta$ measures the difference between the spin-spiral and the ferromagnetic state.

In order to speed up the calculations the force theorem~\cite{Mackintosh:80, Liechtenstein} is used, meaning, the self-consistently obtained charge density of the ferromagnetic state can be used to calculate the properties of any spin-spiral state with a wave vector $\vc{q}$ and then the effect of spin-orbit coupling (SOC) is added in first-order perturbation~\cite{Heide:09} for any state $\vc{q}$. For this step, the number of $\vc{k}$-points was increased to 4096 to improve the accuracy as a small cone angle results in small energy differences. To calculate the inter- and intralayer DM vectors and exchange constants, the total energy with and without SOC was calculated on an $8\times8$ $\vc{q}$-point mesh in the 2D BZ. The real-space parameters were obtained via Fourier-transformation. To obtain accurate model parameters of this system around its ground state (\ie the FM) a small cone angle $\theta$ of 5$\degree$ was used.

\begin{figure}
\centering
\includegraphics[width=\linewidth]{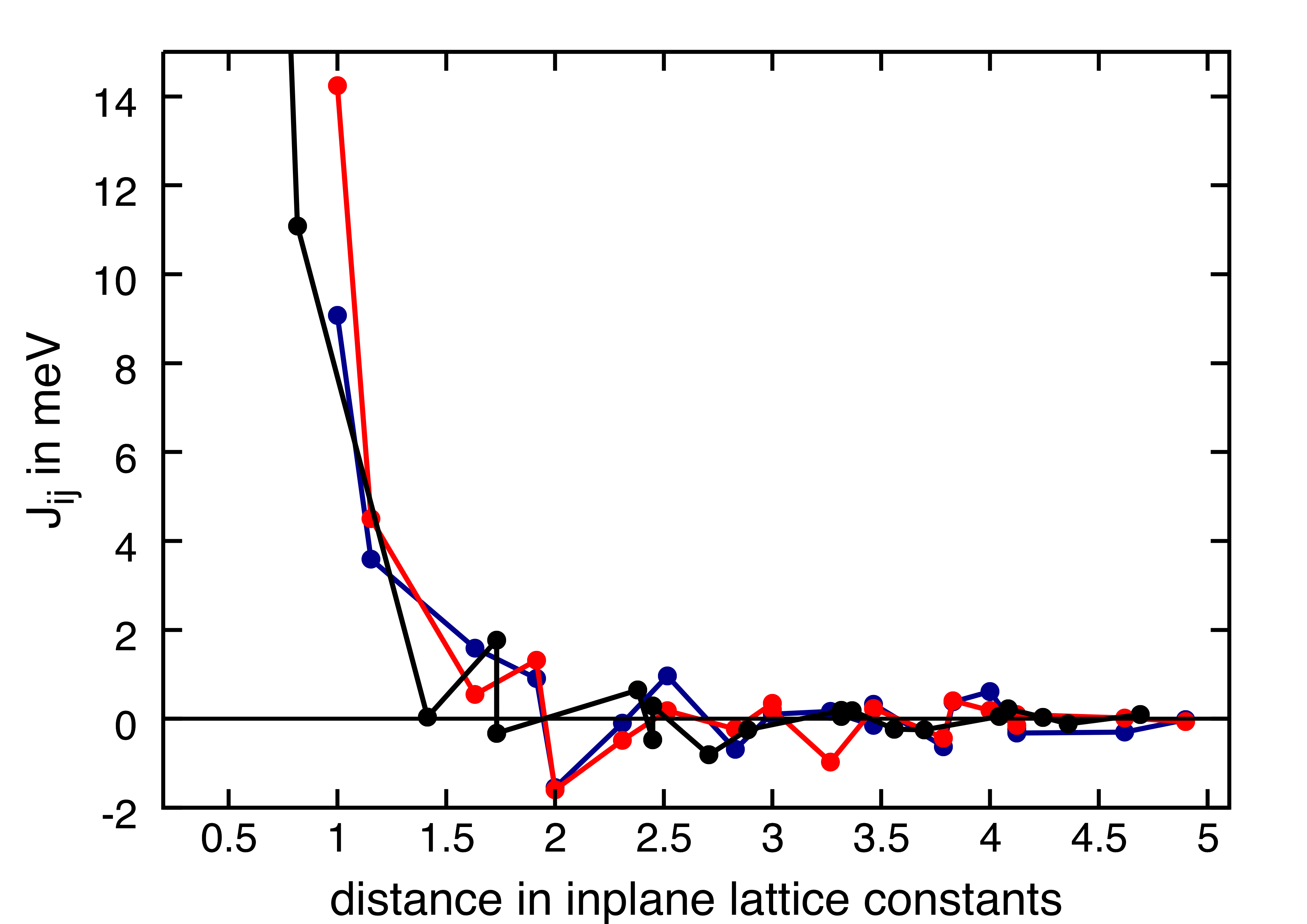}
\caption{\textbf{Exchange parameters $\boldsymbol{J_{ij}}$ in 2Fe/W(110).}  Calculated exchange constants $J_{ij}$ for the intralayer coupling in the surface Fe layer (blue), the intralayer coupling in the interface Fe layer (red) and the interlayer interaction between both layers (black).
}
\label{fig:Jij}
\end{figure}

By fixing the magnetic moments of one of the Fe layers (and the non-magnetic substrate) to ferromagnetic alignment and thus letting the spin-spiral propagate only in the remaining layer we are able to determine the intralayer DM interaction, whereas if the spin-spiral is allowed into both layers simultaneously the interlayer DM interaction can be calculated. The calculated exchange parameters can be found in Supplementary Fig.~\ref{fig:Jij}. We observe the expected decay of $J_{ij}$ with distance $|\vc{R}_i-\vc{R}_j|$, although one should keep in mind that the number of parameters increase linear with distance. On top we find oscillations of size and sign of $J_{ij}$ as function of direction resulting from the very anisotropic 2D Fermi surface.

\section{Supplementary Note 6 $\mid$ Analysis of DM energy contributions to stability of antiskyrmion in 2Fe/W(110)}
To analyse in more detail the role of different DM energy contributions to the stabilization of the antiskyrmion, we plotted in Supplementary Fig.~\ref{fig5} at each site $i$ the local DM energy density by summing over all sites $j$ of the intra- and interlayer DM interactions, \ie within and between the two layers, respectively, evoking the spin-lattice description of the DM energy in Eq.~(1) of the main text with parameters summarized in Fig.~3(d) of the main text and Supplementary Fig.~\ref{fig:Jij}. By inspection of Supplementary Fig.~\ref{fig5}, it becomes obvious, that the main stabilization of the antiskyrmion stems from the contributions within the layers and not between the layers and the dominating contribution comes from the core of the antiskyrmion due to the intralayer DM interaction in the Fe interface layer (see red area in the upper right panel of Supplementary Fig.~\ref{fig5}). However, the rotational sense of the magnetization texture along the [$001$] (\ie y-) direction of the lattice (the spins positioned in Supplementary Fig.~\ref{fig5} \ ``north'' and ``south'' of the center of the antiskyrmion) is energietically disfavoured by the underlying magnetic interaction.  This energy loss is nevertheless compensated by the energy gain of the surface layer along those directions and thus in total, the antiskyrmion gains energy along all crystallographic directions.\par

\begin{figure}[htb]
\centering
\includegraphics[width=\linewidth]{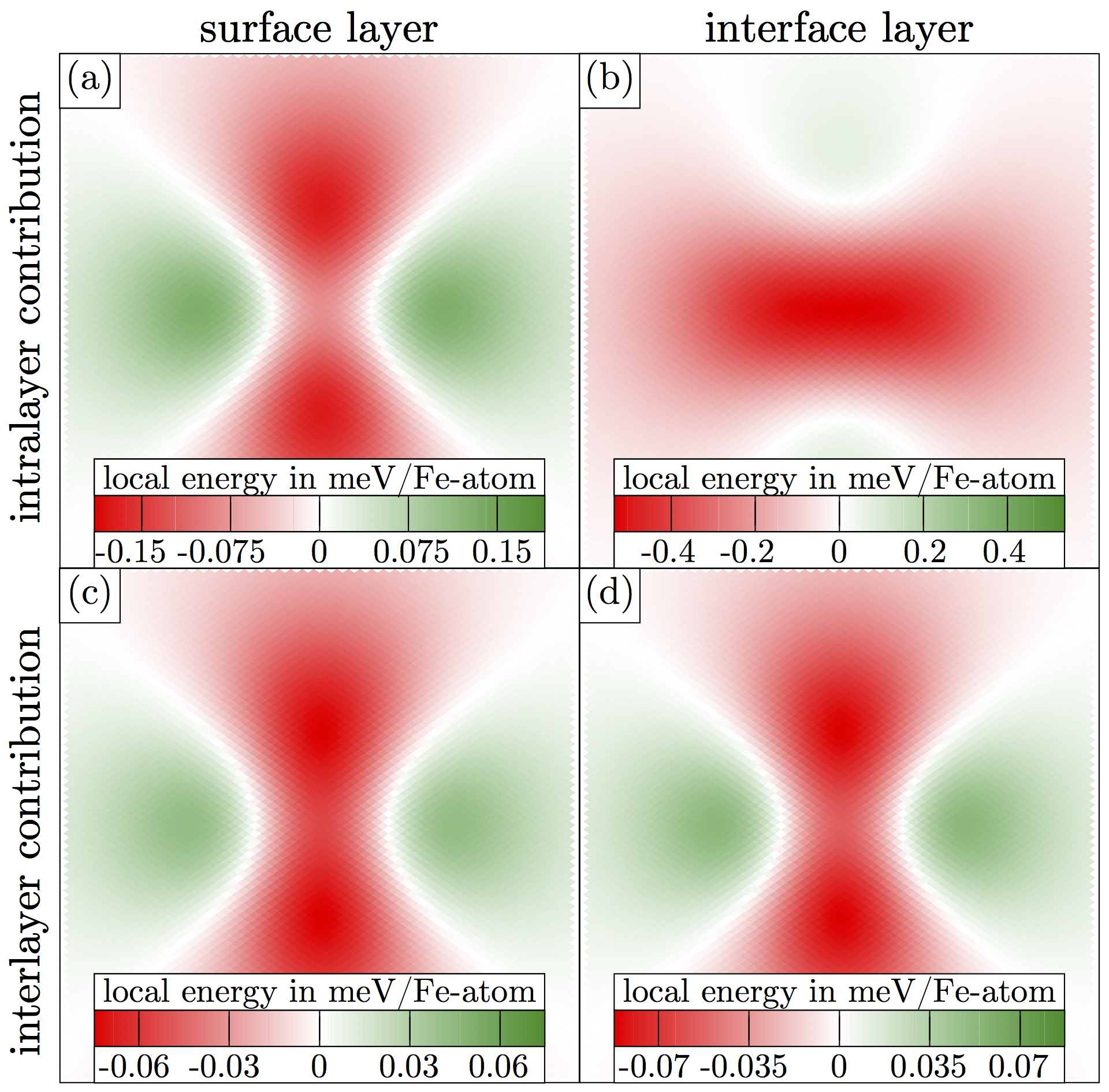}
\caption{\textbf{Local energy distributions of the DMI in 2Fe/W(110).} Shown are the intra- and interlayer contributions of DMI of the Fe surface and interface layer in 2Fe/W(110) for the antiskyrmion shown in Fig.~4 of the main article. A {negative} (red) ({positive} (green)) energy density indicates a local rotational sense that is energetically favoured (disfavoured). An area of $20\mathrm{~nm}\times20\mathrm{~nm}$ around the center of the antiskyrmion is shown. Notice the different colour scales of the different panels.}
\label{fig5}
\end{figure}

As already noticed in Fig.~3 of the main text, the rotational sense enforcing component of the DMI, $(\vc{D}_{ij})_{\varphi}$, of the Fe layer showed changing signs for different neighbours and was mainly satisfied with the formation of the antiskyrmion. Therefore, we carried out spin dynamics simulations where we only included the intralayer DM interactions of the interface layer to check whether then an antiskyrmion could indeed be stable with those parameters. However, we found this is not the case. The antiskyrmion spreads along the y-axis (in agreement with the previously shown analysis) and thus results into a stripe domain wall. On the other hand we realized that the interlayer contributions of the DMI are larger than the interlayer contributions and indeed if we switch off in the spin-dynamics calculation the interlayer contribution, the antiskyrmion is stable. This shows, that for 2Fe/W(110) in particular the interplay of all intralayer DM interactions are crucial for the possible formation of stable antiskyrmions in this system.\par

\section{Supplementary Note 7 $\mid$ Dzyaloshinskii-Moriya vectors for 2Fe/W(110) from the Korringa-Kohn-Rostoker Green-function method}

\begin{table}[tb]\centering
  \begin{ruledtabular}
    \caption{\textbf{Comparison of Results between KKR-GF and FLAPW method}  Microscopic Dzyaloshinskii-Moriya vectors $\vc{D}_{ij}$ calculated with two different methods for a representative pair of atoms for the first three shells within the surface layer (S), the interface layer (I), or between both (IS). The real-space vectors connecting the atoms, $\vc{R}_{ij}$, are given in units of the W-bulk lattice constant $a_{\mathrm{3D}}=$~0.316 nm, and $d = 0.54152\, a_{\mathrm{3D}}$ is the in\-ter\-layer distance.}
    
	\ra{1.2}
    \begin{tabular}{@{}ccccc}
                     &     $\vc{R}_{ij}$   &   \multicolumn{2}{c}{$\vc{D}_{ij}$ [meV]} \\
                     &                         &      KKR         &  FLAPW  \\
\cmidrule[0.4pt](r{0.125em}){1-2}%
\cmidrule[0.4pt](l{0.125em}){3-4}%
S   & $( \frac{1}{\sqrt{2}},\frac{1}{2},    0 )$ & $( -0.389,  1.064,  0     )$ & $( -0.420,  1.062,  0  )$\\
                & $( 0,                 1,              0 )$ & $(  0.159,  0,      0     )$ & $( 0.082,  0,  0 )$\\
                & $( \sqrt{2},          0,              0 )$ & $(  0,     -0.177,  0     )$ & $( 0, 0.005, 0 )$ \\
I & $( \frac{1}{\sqrt{2}},\frac{1}{2},    0 )$ & $(  0.837, -0.639,  0     )$ & $( 0.769,  -0.732, 0 )$\\
                & $( 0,                 1,              0 )$ & $( -0.632,  0,      0     )$ & $( -0.604, 0, 0 )$\\
                & $( \sqrt{2},          0,              0 )$ & $(  0,     -0.270,  0     )$ & $( 0, -0.220, 0 )$\\
IS      & $( 0,                 \frac{1}{2},    d )$ & $( -0.044,  0,      0     )$ & $( 0.013, 0, 0 )$\\    
                & $( \frac{1}{\sqrt{2}},0,              d )$ & $(  0,      0.059,  0     )$ & $( 0, 0.116, 0 )$\\    
                & $( \frac{1}{\sqrt{2}},1,              d )$ & $( -0.283,  0.018, -0.134 )$ & $( -0.336, 0.044, -0.154 )$
	\end{tabular}
	    \label{Tab:comparison_Dij_KKR_FLAPW}
\end{ruledtabular}
\end{table}

Since the FLAPW approach to calculate microscopic atom-pair dependent DM parameters $\vc{D}_{ij}$ used in the main text was newly developed and used here for the first time, for comparison we list in Supplementary Table~\ref{Tab:comparison_Dij_KKR_FLAPW} in addition the $\vc{D}_{ij}$ vectors calculated by means of the Korringa-Kohn-Rostoker Green-function (KKR-GF) method \cite{Bauer_PhDthesis}. Using tke KKR-GF method we first converge the DFT potential including spin-orbit coupling self-consistently, with a chosen magnetization along the $z$-direction, and then perform three one-shot calculations with magnetizations along the $x$, $y$ and $z$ axes, employing the relativistically generalized version~\cite{Udvardi,Ebert} of the method of infinitesimal rotations \cite{Liechtenstein}, in order to access all three components of the DM vector.

The KKR-GF and FLAPW calculations are performed with the same structural parameters and the same exchange-correlation functional \cite{MJW}. The numerical parameters chosen include an angular-momentum cutoff $\ell_\mathrm{max}=3$, a grid of 50 points for integrations along a complex-energy contour with a Fermi smearing of 473~K and 1600 (6400) $k$ points in the full 2D BZ for the integration of the Matsubara pole closest to the real energy axis for self-consistency (in the one-shot calculations employing infinitesimal rotations). 

The Supplementary Table~\ref{Tab:comparison_Dij_KKR_FLAPW} reveals an overall good agreement at the sub-meV level between the DM vectors calcuated by KKR and FLAPW. Test simulations comparing the different sets of DM vectors and exchange parameters disclose the same picture of stable antiskyrmions although the exact diameter of the antiskyrmion and/or the required magnetic field may vary slightly. This is consistent with our finite sub-meV resolution and the number of shells we take into account in our calculations.

A detailed discussion of the two methods employed to extract the $\vc{D}_{ij}$ vectors within the different formalisms, \ie KKR-GF and FLAPW, and a comparison of the results will be published elsewhere.